\documentclass[%
 reprint,
 amsmath,amssymb,
 aps,
floatfix,
]{revtex4-2}

\usepackage{graphicx}
\usepackage{dcolumn}
\usepackage{siunitx}
\usepackage{bm}
\usepackage[colorlinks=true,linkcolor=blue,citecolor=blue]{hyperref}
\usepackage{glossaries}
\usepackage{algorithm}
\usepackage{algpseudocode}
\usepackage{amsmath}
\usepackage{upgreek}
\usepackage[export]{adjustbox}

\let\vec\mathbf

\begin{document}

\preprint{APS/123-QED}

\title{Attaining a strong-field QED signal at laser-electron colliders with optimized focusing}

\author{Christoffer Olofsson}
\affiliation{Department of Physics, University of Gothenburg, SE-41296 Gothenburg, Sweden}
\author{Arkady Gonoskov}%
\affiliation{Department of Physics, University of Gothenburg, SE-41296 Gothenburg, Sweden}

\date{\today}

\begin{abstract}

Colliding bunches of high-energy electrons with intense laser pulses provides a basis for studying strong-field QED processes enabled by high values of quantum non-linearity parameter $\chi$. Nevertheless, the signal deconvolution is intricate due to probabilistic nature of the processes and shot-to-shot variation of the impact parameter, which disfavors the use of tight focusing. We propose a concept for distinguishing the signal of high-$\chi$ emissions that enables the use of optimal focusing to attain the highest $\chi \approx 5.25 \left(\varepsilon / (\text{1~GeV})\right)\left(P/(\text{1~PW})\right)^{1/2}\left((\text{1~}\mu\text{m})/\lambda\right)$ for a given electron energy $\varepsilon$, laser power $P$ and wavelength $\lambda$. Reaching such $\chi$ with f/2 focusing requires more than 10 times higher power.
\end{abstract}

\maketitle

\section{Introduction}

High-intensity laser facilities \cite{eli, xcels, kitagawa.qe.2004, kawanaka.jpcs.2016, danson.hplse.2019} in combination with conventional or laser-based electron accelerators open up opportunities to study extreme regimes of radiation reaction (RR) and of other effects due to strong-field quantum electrodynamics (SFQED) \cite{dipiazza.rmp.2012, gonoskov.rmp.2022, fedotov.arxiv.2022}. The effect of laser-generated electromagnetic fields on an electron is characterized by the dimensionless acceleration in its rest frame $\chi = \gamma E_\text{cr}^{-1} \left(\left(\vec{E} + \left(\vec{v}/c\right) \times \vec{B}\right)^2 - \left(\vec{E} \cdot \vec{v} /c\right)^2\right)^{1/2}$, where $\vec{v}$ and $\gamma$ are the electron velocity and gamma factor, $c$ is the speed of light, $\vec{E}$ and $\vec{B}$ are the electric and magnetic field vectors. The critical field of QED is defined as $E_\text{cr} = m^2 c^3 /e\hbar$, where $\hbar$ is the reduced Planck constant, $m$ and $e$ are the electron mass and charge. The experimental objectives range from testing existing predictions at $\chi \gtrsim 1$, which designates the quantum regime of RR, to the detection of early signs of unknown behaviour at $\chi \gtrsim 1600$, which  demarcates qualitatively unexplored regimes characterized by the conjectured breakdown of perturbative nonlinear QED \cite{ritus.ap.1972, narozhny.prd.1980, fedotov.jpcs.2017}. 

In the '90s the collision of 46.6-GeV electrons with focused laser pulses was used to observe multiphoton Compton scattering \cite{bula.prl.1996} and multiphoton Breit-Wheeler pair creation \cite{burke.prl.1997}. Revisiting this experimental configuration with extended experimental program is a matter of several initiatives, including the E320 collaboration at FACET-II \cite{yakimenko.prab.2019} and the LUXE (Light Und XFEL Ex-periment) collaboration at the European XFEL~\cite{abramowicz.arxiv.2019}. Another experimental alternative is based on the replacement of conventional acceleration by laser wake-field acceleration (LWFA). In recent experiments, signatures of RR at $\chi \sim 0.1$ have been observed by colliding LWFA electrons with a laser pulse focused by a f/2 parabolic mirror \cite{cole.prx.2018, poder.prx.2018}. For further studies at higher $\chi$, apart from increasing electron energy and laser power, it is natural to consider how large increase of $\chi$ we can get from more advanced focusing.

It is reasonable to think that this question has little practical meaning because a smaller strong-field region formed by tighter focusing implies an enhanced role of the shot-to-shot variation of the impact parameter, which is present due to limited capabilities for laser-electron beam alignment. In addition, $4\pi$ focusing may require unprecedentedly large parabolic mirrors, while reaching strong fields itself may require low vacuum \cite{gonoskov.prl.2014} to prevent early cascade development \cite{bulanov.prl.2010b}. Finally, it is unclear how to distinguish the signal of SFQED at high $\chi$ from the dominating signal of emissions at low $\chi$.

In this paper we consider the problem of detecting and measuring the evidences of SFQED predictions that have not yet been experimentally verified. Specifically, we elaborate a strategy to reveal such evidences without accessing the regimes when these phenomena become prominent and significantly change the interaction physics. This is done under the assumption that the influence of these phenomena on the interaction process gradually rises with $\chi$. In this case high accuracy of measurements and/or statistical analysis can be used to infer the sought-for signal from experiments with $\chi$ value well below than that needed for the prominent change of physics. One example of a possible objective is to detect the effective mass change discussed in Ref.~\cite{yakimenko.prl.2019}.

As the main result of the paper, we propose a way to extract the signal of SFQED events occurring at high $\chi$ and for electrons having known initial energy, such that high localization of the strong-field region is no longer an obstacle. Assuming this possibility, we determine the optimal focusing geometry, which we call bi-dipole wave, and identify prospects and limitations of the proposed concept.\\

\section{Signal distinguishing}
\begin{figure} 
\includegraphics[width=0.8\columnwidth, scale=1]{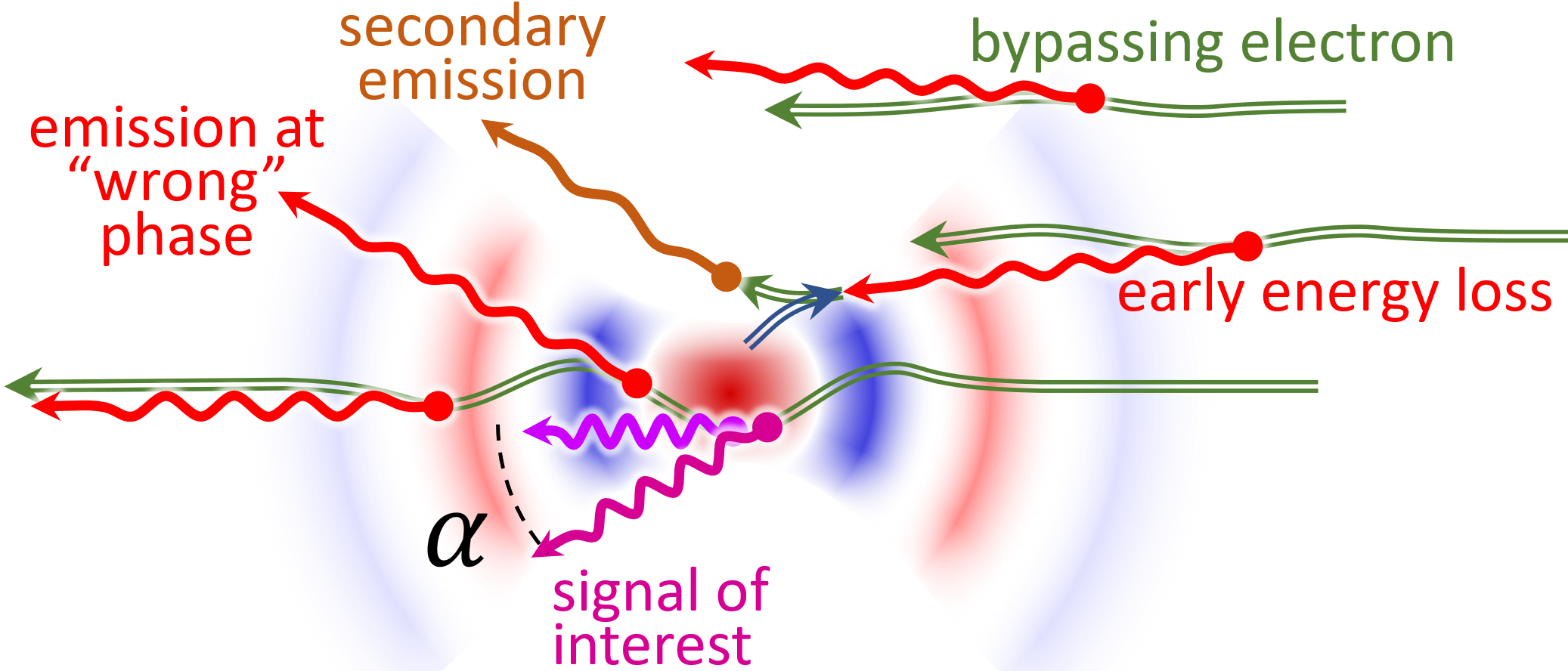}
\caption{Schematic illustration of interaction scenarios for various electrons passing through a tightly focused laser pulse.}
\label{signal}
\end{figure}

Since emissions probabilistically happen at unknown field strength and $\gamma$ (in case of prior emissions), it is not possible to do a direct measurement of SFQED emission rate as a function of $\chi$, fractional photon energy $\hbar\omega/mc^2\gamma$ and other parameters of interest. Instead, we have to make inferences by comparing experimental and numerical results for some measurable quantities/distributions that notably depend on this rate but preferably independent of unmeasurable shot-to-shot variations. The influence of the impact parameter becomes largely eliminated if the electron bunch is underfocused/diverged such that tight laser focusing occurs somewhere within a nearly uniform flux of electrons. Despite yielding smaller number of emissions in the strong-field region, we choose this layout because it corresponds to data accumulation from repeated experiments and, more importantly, permits reaching higher $\chi$ by arbitrary tight focusing. An explicit signal of high-$\chi$ SFQED rate is carried by electrons that emitted a single photon at about maximal $\chi$. Unfortunately, at the detector these electrons can hardly be distinguished from the majority of those bypassing the strong-field region, emitting multiple photons and/or at low $\chi$, and those being generated by the Breit-Wheeler process (see fig.~\ref{signal}). However, a fortunate opportunity to distinguish an informative signal occurs for photons based on their energy and deviation angle $\alpha$, the angle between propagation directions of a photon and initial electrons. Firstly, among electrons having initial energy only the ones passing through the strong-field region have the chance to emit photons with large $\alpha \sim a_\text{max}/\gamma$, where $a_\text{max}$ is the peak field amplitude in units of $mc\omega_0/e$ with $\omega_0$ being the laser frequency. Note that $\alpha$ is required to be measurable but can still be assumed small due to large $\gamma$ so that electrons deviate by a distance negligible as compared to the wavelength $\lambda$ (in fig.~\ref{signal} the deviation is exaggerated for illustrative purposes). Secondly, the higher the energy of a detected photon the less likely it was emitted after another emission or by a newly generated particle. It is clear that among photons with large energy and $\alpha$ we can expect a large part of those carrying the signal. As we show further, this makes possible reaching a given confidence level of statistical inferences with many orders of magnitude smaller number of shots than that required in case of electron-based diagnostics. Finally, note that fig.~\ref{signal} shows an unfavorable phase dependence: $\alpha = 0$ when the field peaks and vice versa. Using circular polarization (CP) makes $\alpha$ and $\chi$ correlated (see fig.~\ref{fig_scheme_b}), permitting direct measurements of the rate as a function of $\chi$.\\

\section{Optimal focusing} The possibility of creating the strongest field for a given power $P$ by the dipole wave \cite{bassett.oa.1986, gonoskov.pra.2012, jeong.oe.2020} using multiple colliding laser pulses (MCLP) \cite{bulanov.prl.2010} has been recognized to enable many possibilities, ranging from particle trapping and photon generation \cite{gonoskov.prl.2014, gonoskov.prx.2017, magnusson.pra.2019, magnusson.prl.2019} to the creation of sustained electromagnetic cascades and extreme electron-positron plasma states \cite{efimenko.pre.2019, efimenko.sr.2018}. Nevertheless, the problem of maximizing $\chi$ for a given power received little attention in the literature. In Ref.~\cite{gelfer.pra.2015} the analysis was restricted to the case of several beams focused to a point from the directions laying in a single plane. Here we consider the general case.

To quantify the role of focusing we notice that in the focal region the intensity is proportional to $P$ and inversely proportional to the focal area being proportional to $\lambda^{2}$. Therefore the field strength scales as $\propto \lambda^{-1} P^{1/2}$ and $\chi \propto \gamma \lambda^{-1} P^{1/2}$. Thus, a focusing geometry can be quantified by a single dimensionless parameter $\kappa$ that determines the peak $\chi$ value (we assume $|\vec{v}| \approx c$, $\gamma \gg a_\text{max}$):
\begin{equation} \label{eq:chi_geom}
    \chi_\text{max} = \kappa \left(\frac{\varepsilon}{\text{1~GeV}}\right)
    \left(\frac{P}{\text{1~PW}}\right)^{1/2}
    \left(\frac{\lambda}{\text{1~}\mu\text{m}}\right)^{-1}.
\end{equation}
The problem of maximizing $\kappa$ belongs to the class of optimization problems in optics. In 1986, using multipole expansion, Bassett showed that the dipole component provides the highest possible energy density for a given $P$ and this optimal component also provides the strongest field strength $a_0 \approx 780 \left(P/(\text{1~PW})\right)^{1/2}$ \cite{bassett.oa.1986}. 

To determine the optimal geometry we assume that the maximal value of $\chi$ is achieved at the origin of spherical coordinate system $(r, \theta, \phi)$ with the electric field pointing towards $\theta = 0$. Assuming that the incoming wave is monochromatic, the field can be expressed as the real part of $\exp(-i\omega_0 t)$ multiplied by some complex field $(\vec{E}^\chi, \vec{B}^\chi)$, which in turn can be expressed using the basis of electric $(\vec{E}^E, \vec{B}^E)$ and magnetic $(\vec{E}^B, \vec{B}^B)$ multipoles (exact solutions of Maxwells' equations) given by \cite{panofsky.2005}:
\begin{align}
\begin{split}
    &E^E_r = l(l+1)r^{-1}j_l(kr)Y^m_l(\theta, \phi),\\
    &E^E_\theta = r^{-1} \partial_r\left(rj_l(kr)\right)
    \partial_\theta Y^m_l(\theta, \phi),\\
    &E^E_\phi = im\sin^{-1}\theta r^{-1} \partial_r\left(rj_l(kr)\right) Y^m_l(\theta, \phi),\\
    &B^E_r = 0,\:\:
    B^E_\theta = km\sin^{-1}\theta j_l(kr)Y^m_l(\theta, \phi),\\
    &B^E_\phi = i k j_l(kr) \partial_\theta \left(Y^m_l(\theta, \phi) \right),\\
    &\vec{E}^B = -\vec{B}^E, \:\: \vec{B}^B = \vec{E}^E,
\end{split}
\end{align}
where $l = 1, 2, 3, ...$, $m = -l, -l+1, ... l$, $k = \omega_0/c$, $j_l(kr)$ and $Y^m_l(\theta, \phi)$ are spherical Bessel functions and spherical harmonics, respectively. The subscripts denote vector components along the unit vectors.

Bassett showed that in terms of incoming power the multipolar components are additive, i.e. for any their combination the incoming power is the sum of incoming powers of the components (see Sec.~3 in Ref.~\cite{bassett.oa.1986}). Following Bassett we consider the limit $r \rightarrow 0$ and notice that only six components contribute to the field at $r = 0$ ($\vec{E}$ is formed by $\vec{E}^E_{l,m}$ and $\vec{B}$ is formed by $\vec{B}^B_{l,m}$, in both cases $l = 1, m = -1, 0, 1$):
\begin{align}
    &\vec{E}^E_{1,-1} = \vec{B}^B_{1,-1} = k\left(6\pi\right)^{-1/2}\left(\hat{\vec{x}} - i \hat{\vec{y}}\right),\\
    &\vec{E}^E_{1,0} = \vec{B}^B_{1,0} = k\left(3\pi\right)^{-1/2}\hat{\vec{z}},\\
    &\vec{E}^E_{1,1} = \vec{B}^B_{1,1} = -k\left(6\pi\right)^{-1/2}\left(\hat{\vec{x}} + i \hat{\vec{y}}\right),
\end{align}
where $\hat{\vec{x}}$, $\hat{\vec{y}}$ and $\hat{\vec{z}}$ are the Cartesian system unit vectors pointing towards $(\theta = \pi/2, \phi = 0)$, $(\theta = \pi/2, \phi = \pi/2)$ and $(\theta = 0)$, respectively. According to our assumption $\vec{E}^\chi$ is pointing towards $\theta = 0$ and thus it is formed exclusively by the component $\vec{E}^E_{1,0}$. Without loss of generality we can assume that the coordinate system is oriented so that $\vec{B}^\chi(r = 0)$ is laying in the $xz$ plane and thus it is formed by a combination of $\vec{B}^B_{1,0}$ and $2^{1/2}\left(\vec{B}^B_{1,-1} - \vec{B}^B_{1,1}\right)$ components (the factor is chosen to provide synchronous peaking). The components $\vec{E}^E_{1,0}$ and $\vec{B}^E_{1,0}$ correspond to the electric and magnetic dipole waves with symmetry axis along $\hat{\vec{z}}$, whereas $2^{1/2}\left(\vec{B}^B_{1,-1} - \vec{B}^B_{1,1}\right)$ corresponds to the magnetic dipole wave with symmetry axis along $\hat{\vec{x}}$. Given that the power of components is additive, we can describe all cases by splitting the total power $P$ into three portions: $aP$ is delivered by the electric dipole wave, whereas the portions $bP$ and $(1 - a - b)P$ are delivered by the magnetic dipole waves with symmetry axes along $\hat{\vec{z}}$ and $\hat{\vec{x}}$, respectively; $0 \leq a, b \leq 1, a + b \leq 1$. The strength of components in relativistic units is given by 
\begin{align}
    &\vec{E} = a_d \left(P/(\text{1~PW})\right)^{1/2} a^{1/2} \hat{\vec{z}}, \\
    &\vec{B} = a_d \left(P/(\text{1~PW})\right)^{1/2}\left(b^{1/2} \hat{\vec{z}} + \left(1 - a - b\right)^{1/2} \hat{\vec{x}}\right),
\end{align}
where $a_d \approx 780$. If we were interested in the maximal energy density ($\left(E^2+B^2\right)/8\pi$) all the cases were indifferent because the energy density is independent of $a$ and $b$. Nevertheless, searching for the strongest possible acceleration yields one specific optimum. First we note that the maximal Lorentz force is achieved if the electron propagates along the $y$ axis. In this case the absolute value of the Lorentz force and the $\chi$ value is proportional to 
\begin{equation}
    \chi \propto \left(\left(a^{1/2} + \left(1 - a - b\right)^{1/2}\right)^2 + b\right)^{1/2}.
\end{equation}
Searching for the maximum of $\chi^2$ we first notice that $\partial \chi^2 /\partial b \leq 0$ for all $a$ and $b$, meaning that the maximum is achieved at $b = 0$. Next, we compute $\partial \chi^2 /\partial a$ and determine that the maximum is achieved at $a = 1/2$. 

As one can see the maximum corresponds to the equal destitution of energy between electric and magnetic dipole waves that have perpendicular axes. That is why we choose to call this geometry bi-dipole wave. The field strength is $|\vec{E}^\chi| = |\vec{B}^\chi| = a_\text{max} = a_d \left(P/(\text{1~PW})\right)^{1/2} /\sqrt{2} \approx 550 \left(P/(\text{1~PW})\right)^{1/2}$ (in relativistic units) and the value of $\kappa$ is approximately 5.25. 

So far we were considering monochromatic radiation. Lifting this limitation, i.e. considering electromagnetic pulse shape as another matter of optimization, yields an ill-posed maximization problem: higher frequency gives smaller focal volume and thereby higher field strength for the same power, which means that $\chi$ is unbound from above in case of unrestricted spectrum. This is also manifested by the $\chi_\text{max}$ dependence on the wavelength shown in eq.~(\ref{eq:chi_geom}). Nevertheless, a fortunate possibility to assess practically important pulsed solutions is provided by the theory of dipole pulses developed in Ref.~\cite{gonoskov.pra.2012}. Since the bi-dipole wave is a sum of two dipole waves, we can generalize our result and define a bi-dipole pulse as a sum of two dipole pulses (one electric and one magnetic, both synchronized in time, with perpendicular axes of symmetry). For example, using this we can answer one -- probably quite practical -- question: what benefit can we get from the shortness of a focused Gaussian-like pulse? Using the theory of dipole pulses we obtain that the relative increase of $\chi$ is $3 \ln 2 \tau_0^{-2}/\pi^2$, where $\tau_0$ is the pulse duration in cycles, determined according to the full-width half-maximum for intensity. Practically this means that the benefit is moderate: even for 2-cycle pulse it is about 5\%, while for common Ti-Sapphire pulses ($\sim$ 5 cycles) it is $< 1$~\%.\\

\section{Bi-dipole wave structure}

We continue by considering the structure of the bi-dipole wave and the way to generate it in practice. To within a constant factor the electric field vector in the far field region can be given by:

\begin{figure*}
\includegraphics[width=\textwidth, scale=0.3]{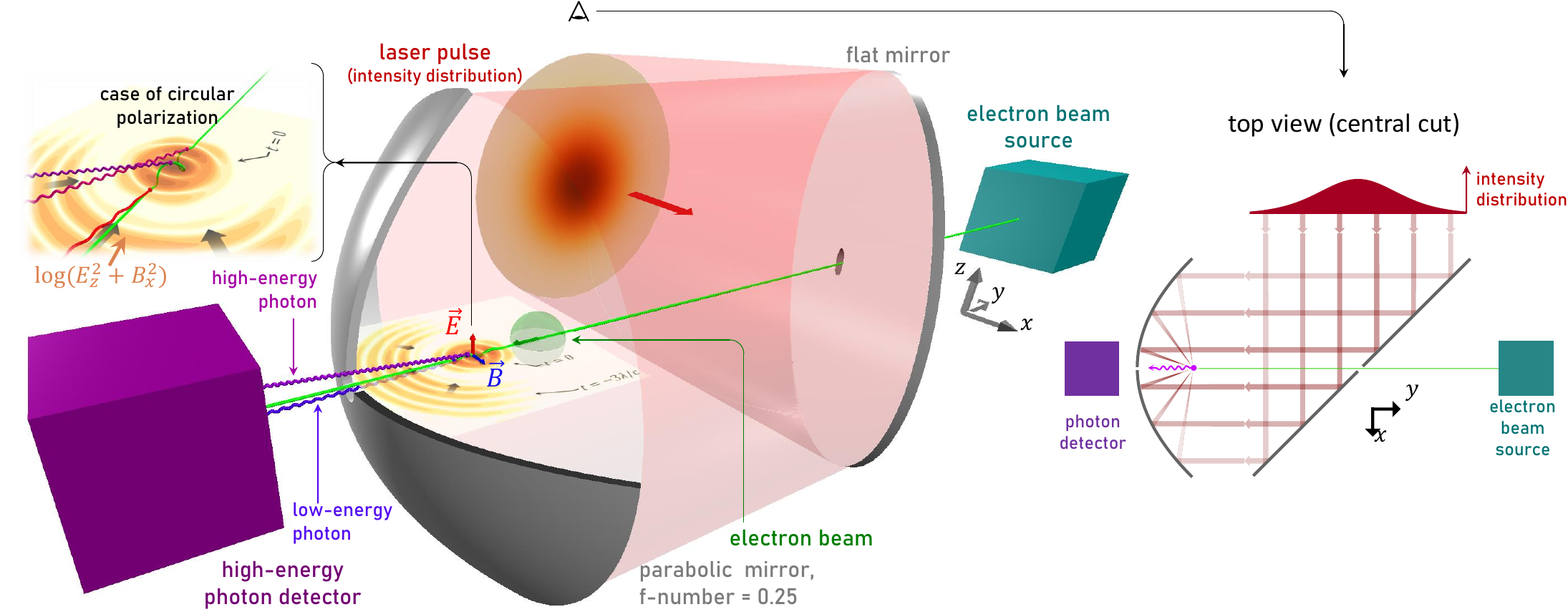}
\caption{Schematic representation of a possible experimental setup with an insert showing the case of circular polarization (deviation angles are exaggerated).}
\label{fig_scheme_a}
\end{figure*}

\begin{figure} 
\includegraphics[width=0.8\columnwidth, scale=1]{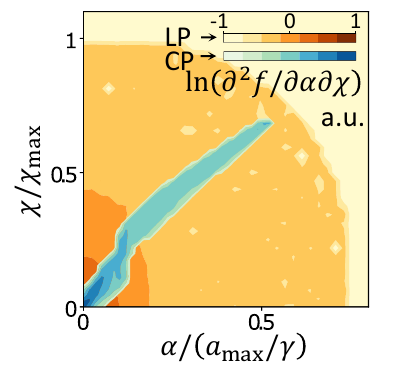}
\caption{Relative time that the electrons spend with various $\alpha$ and $\chi$ while passing through the dipole wave for CP and LP cases shown in fig.~\ref{fig_scheme_a}.}
\label{fig_scheme_b}
\end{figure}

\begin{align}
    &\vec{E}(r \rightarrow \infty) \propto r^{-1}\left(\vec{E}^e + \vec{E}^b\right),\\
    &\vec{E}^e = \left(\hat{\vec{z}} \times \vec{n}\right) \times \vec{n},\:\:
    \vec{E}^b = \left(\hat{\vec{x}}\times \vec{n}\right),
\end{align}
where $\vec{n} = \vec{r}/r$, whereas $\vec{E}^e$ and $\vec{E}^b$ are proportional to the electric field vectors of the radiation forming electric and magnetic dipole waves, respectively. The signs are chosen so that the constructive summation of the electric and magnetic components is provided for a charge propagating towards the negative $y$ direction.

Along each direction the dipole waves are formed by linearly polarized, syn-phased waves and thus the bi-dipole wave is also formed by linearly polarized wave. Let us demonstrate that the distribution of intensity $I$ of this wave is symmetric about $y$ axis. To do so we compute the intensity as a function of $\vec{n}$:
\begin{align*}
\begin{split}
    &\vec{E}^e = n_x n_z \hat{\vec{x}} - \left(n_x^2 + n_y^2\right)\hat{\vec{z}} + n_y n_z \hat{\vec{y}},\:\:
    \vec{E}^b = n_y \hat{\vec{z}} - n_z \hat{\vec{y}}, \\
    &I \propto \left|\vec{E}^e + \vec{E}^b\right|^2 = n_x^2 n_z^2 + \left(n_x^2 + n_y^2 - n_y\right)^2 + n_z^2 \left(n_y - 1\right)^2. 
\end{split}
\end{align*}
To demonstrate the axial symmetry of $I$ we introduce a spherical coordinate system $(\theta_y, \varphi)$ so that the unit vector has the components:
\begin{align*}
\begin{split}
&n_y = \cos\theta_y\\
&n_x = \sin\theta_y\cos\varphi \\
&n_z = \sin\theta_y\sin\varphi
\end{split}
\end{align*}
Given that $\partial n_y / \partial\varphi = 0$, $\partial n_x /\partial\phi = - n_z$, $\partial n_z /\partial\varphi = n_x$, we can compute
\begin{equation}
    \frac{\partial I}{\partial \varphi} \propto 2 n_z n_x \left(1 - n_z^2 - n_y^2 - n_z^2\right) = 0,
\end{equation}
which proves the axial symmetry.

If $\varphi = 0$ is chosen, s.t. $\mathrm{cos}(\varphi)=1$ and $\mathrm{sin}(\varphi)=0$, then in the expression for $I$ only the middle term remains and 
\begin{equation} \label{eq:intensity_scale}
    I \propto (1-\cos\theta_y)^2 /r^2.
\end{equation}
As we can see the radiation is arriving predominantly from the negative $y$ hemisphere. 

One possible way to form the bi-dipole wave is the reflection of an appropriate laser beam propagating towards negative $y$ direction from a parabolic mirror. Let us compute the polarization and intensity distribution in such a beam. During reflection the electric field component along the normal $\vec{N}$ to the mirror is reversed. Thus, to within a factor the electric field before reflection is given by $\vec{E}^p = \vec{E} - 2 (\vec{E} \cdot \vec{N})\vec{N}$, where the normal can be expressed as $\vec{N} = \left(\hat{\vec{y}} - \vec{n}\right)/\left|\hat{\vec{y}} - \vec{n}\right|$. We note that $(\vec{E}\cdot\vec{N}) = n_z(n_y - 1)/\left|\hat{\vec{y}} - \vec{n}\right|$ and compute the $x$ component of $\vec{E}^p$:
\begin{align*}
\begin{split}
    E^p_x &= n_x n_z - 2n_z\left(n_y - 1\right)\left(-n_x\right)\left(n_x^2 + n_z^2 + \left(n_y - 1\right)^2\right)^{-1}\\
    &=  n_x n_z + 2 n_z (n_y - 1) n_x\left(n_x^2 + n_z^2 + \left(n_y - 1\right)^2\right)^{-1}\\
    & = 0.
\end{split}
\end{align*}
As we can see the beam to be reflected has linear polarization exactly along $z$ axis everywhere. Due to such a fortunate property, this configuration has been considered by Sheppard and Larkin \cite{sheppard.jmo.1994} as a notably practical option among all mixed dipole waves that yield the highest electromagnetic field density under focusing of a given power. As we demonstrated, exactly this option also gives the highest value of $\chi$. 
Using eq.~(\ref{eq:intensity_scale}) one can compute the intensity at the mirror and, using local $\vec{N}$, the intensity distribution in the beam to be reflected:
\begin{equation}
    I^p(R) \propto \left(\left(R/2L\right)^2 + 1\right)^{-4},
\label{intensityShape}
\end{equation}
where $L$ is the distance to the mirror and $R$ is the distance to the $z$ axis in transverse plane. The resultant scheme of potential experiments is illustrated in fig.~\ref{fig_scheme_a}.

\begin{figure} 
\includegraphics[width=0.8\columnwidth, scale=1]{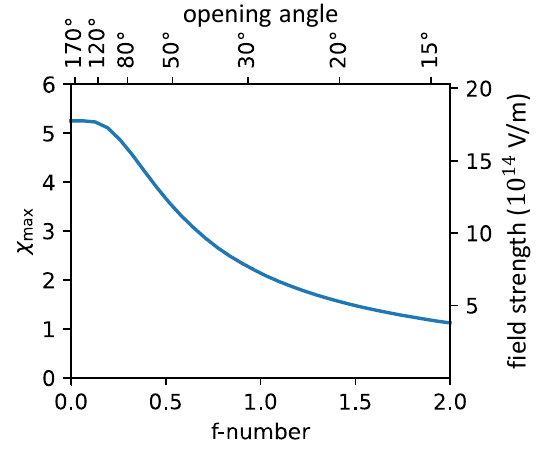}
\caption{The dependency of $\chi_\text{max}$ on the f-number that characterizes the radius of parabolic mirror show in fig.~\ref{fig_scheme_a}. The axis above shows corresponding values of the opening angle. The shown values are numerically obtained for $P = 1$~PW, $\varepsilon = 1$~GeV and $\lambda = 1$~$\mu$m, whereas the values for other cases can be obtained using scaling $\chi_\text{max} \propto \varepsilon \lambda^{-1} P^{1/2}$. The axis on the right shows the corresponding peak values of electric/magnetic field strength that scales as $\propto \lambda^{-1} P^{1/2}$.}
\label{fig_scheme_c}
\end{figure}

The infinite parabolic mirror that forms the bi-dipole wave by reflecting an intensity-shaped beam has to be limited in practice. To facilitate more practical consideration of the use of bi-dipole wave in experiments we consider how the peak value of $\chi$ depends on the parabolic mirror radius $R_\text{max}$. The limitation $R < R_\text{max}$ can be expressed via the f-number$=L/2R_\text{max}$ or via the opening angle $\zeta = \pi/2 + \arctan{\left(R_\text{max}/4L - L/R_\text{max}\right)}$ of the cone that encompasses the rays coming to the focus. In fig.~\ref{fig_scheme_c} we show numerically computed peak value $\chi_\text{max}$ as a function of f-number. We assume the shape given by eq.~(\ref{intensityShape}) with restriction $R < R_\text{max}$ and keep the total incoming power $P = 1$~PW in all cases. We assume that $\varepsilon = 1$~GeV and $\lambda = 1$~$\mu$m, whereas the values for other cases can be obtained using scaling $\chi_\text{max} \propto \varepsilon \lambda^{-1} P^{1/2}$. The right axis shows the corresponding peak values of electric/magnetic field strength that scales as $\propto \lambda^{-1} P^{1/2}$.\\

\section{Numerical analysis}

\begin{figure}[htb] 
\hspace*{0.4cm}
\includegraphics[center]{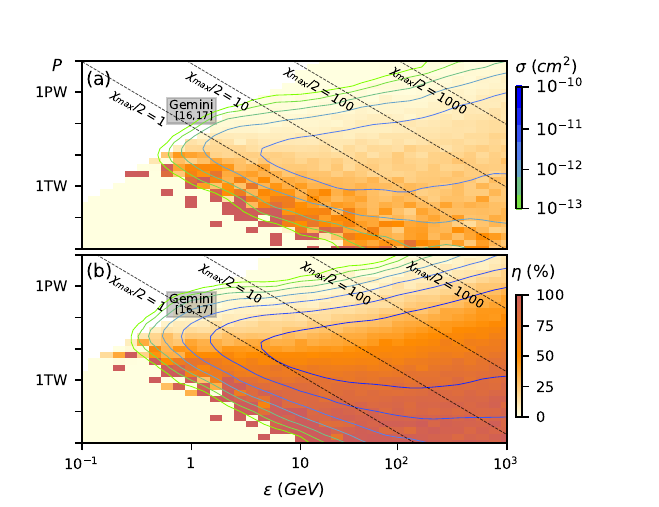}
\caption{
Effective cross-section $\sigma$ and signal ratio $\eta$ computed numerically for linear (a) and (b) circular polarization.
}
\label{fig:pscan}
\vspace*{-0.3cm}
\end{figure}

Having determined the optimal focusing, we can assess the concept capabilities, which we characterize by a signal ratio $\eta = N_\text{signal}/N_\text{total}$ and an effective cross-section $\sigma$. Here $N_\text{total}$ is the number of photons detected with $\alpha > 0.6 a_\text{max}/\gamma$ and $\hbar\omega > mc^2\gamma/2$, whereas among them $N_\text{signal}$ photons are emitted at $\chi > \chi_\text{max}/2$ and by electrons that hadn't lost more than 1\% of initial energy prior to the emission. The cross-section is defined as $\sigma = N_{\text{signal}} / n_e \tau_l c$, where $n_e$ is the density of streaming electrons and $\tau_l$ is the laser pulse duration. In fig.~\ref{fig:pscan} we demonstrate the results of simulations for various $P$ and $\varepsilon$ values that can be relevant to current and upcoming experimental capabilities. We consider a laser pulse that has a Gaussian profile with a duration of 5 cycles (FWHM for intensity); $\lambda= 0.8$~$\mu$m. In case of CP the peak amplitude is $a_\text{max}^\text{CP} = a_\text{max}/\sqrt{2}$ and we use a modified selection rule $\alpha > 0.7 a_\text{max}^\text{CP}/\gamma$.

To exemplify the results, let us estimate the number of shots needed to reach 3-$\sigma$ confidence level for detecting a 1\% deviation of the rate at $\chi > 10$. Assuming current progress on LWFA \cite{gonsalves.prl.2019}, we consider a 10~GeV electron bunch of total charge 100 pC ($6 \times 10^8$ electrons) that is spread over a spherical volume with $5$~$\mu$m diameter, so that $n_e \approx 10^{19}$~cm$^{-3}$. Using $N$ shots and cumulative $N_\text{total}$ as a test statistic means that $3\sigma_N \leq 0.01 N_\text{signal}$, where the variance of $N_\text{total}$ is $\sigma_N^2 \approx N_\text{total}$. From Fig.~\ref{fig:pscan} we see that for $\varepsilon = 10$~GeV the value $\chi_\text{max}/2 = 10$ is achieved at $P \approx 100$~TW with $\sigma \sim 10^{-11}$~cm$^2$ and $\eta \sim 0.1$. Substituting these values, we estimate that we have $\sim 5 \times 10^4$ photons per shot, while we need $\sim (3/0.01\eta)^2 \approx 10^7$ photons, and thus $N \sim 200$ shots ($N \propto 1/\sigma \eta ^{2}$).

Let us compare this to the number of shots required in case of using electron energy loss for the test statistic. Assuming best case scenario, we consider all the electrons passing through a 5-cycle laser pulse with $a_\text{max} \approx 200$ to reach $\chi = 10$. At $\chi \sim 10$ the mean free path is $\approx 15 \lambda \chi^{1/3} / a_\text{max}  \approx 0.15 \lambda$ \cite{gonoskov.rmp.2022} and electrons need to propagate about $2\lambda$ to reach $\chi = 10$. Thus, only a fraction of $\eta \sim \exp(-2/0.15) \approx 10^{-6}$ of initial electrons can keep high $\gamma$ to be affected by the deviation at $\chi > 10$. This means that we need cumulatively $\sim (3/0.01\eta)^2\approx 10^{17}$ electrons and having $6 \times 10^8$ electrons per shot this requires $\sim 10^8$ shots.\\

\section{Conclusions}

In conclusion, we showed that for laser-electron colliders the energy and deviation angle of emitted photons can be used to fetch out the characteristics of high-$\chi$ SFQED rates despite the background of low-$\chi$ emissions. This permits using tight focusing to boost $\chi$ values, for which we determined the limit given by the so-called bi-dipole wave. The concept prospects were characterized by the effective cross-section and signal ratio, indicating that PW-10GeV-class facilities can study $\chi \sim 10-100$, while $\chi \sim 10^3$ requires higher energy and/or power in combination with cascade suppression. The latter encourages further studies on electron injection \cite{magnusson.prl.2019, blackburn.njp.2019}, as well as 
on the use of shorter pulses \cite{rivas.sr.2017} or pulse steepening \cite{gustafson.pr.1969, kaw.pf.1970, gonoskov.prl.2009, reed.apl.2009, wang.prl.2011}.\\

The authors acknowledge support from the Swedish Research Council (Grant No. 2017-05148) as well as computational resources provided by the Swedish National Infrastructure for Computing (SNIC).

\bibliography{literature}

\begin{thebibliography}{40}%
\makeatletter
\providecommand \@ifxundefined [1]{%
 \@ifx{#1\undefined}
}%
\providecommand \@ifnum [1]{%
 \ifnum #1\expandafter \@firstoftwo
 \else \expandafter \@secondoftwo
 \fi
}%
\providecommand \@ifx [1]{%
 \ifx #1\expandafter \@firstoftwo
 \else \expandafter \@secondoftwo
 \fi
}%
\providecommand \natexlab [1]{#1}%
\providecommand \enquote  [1]{``#1''}%
\providecommand \bibnamefont  [1]{#1}%
\providecommand \bibfnamefont [1]{#1}%
\providecommand \citenamefont [1]{#1}%
\providecommand \href@noop [0]{\@secondoftwo}%
\providecommand \href [0]{\begingroup \@sanitize@url \@href}%
\providecommand \@href[1]{\@@startlink{#1}\@@href}%
\providecommand \@@href[1]{\endgroup#1\@@endlink}%
\providecommand \@sanitize@url [0]{\catcode `\\12\catcode `\$12\catcode
  `\&12\catcode `\#12\catcode `\^12\catcode `\_12\catcode `\%12\relax}%
\providecommand \@@startlink[1]{}%
\providecommand \@@endlink[0]{}%
\providecommand \url  [0]{\begingroup\@sanitize@url \@url }%
\providecommand \@url [1]{\endgroup\@href {#1}{\urlprefix }}%
\providecommand \urlprefix  [0]{URL }%
\providecommand \Eprint [0]{\href }%
\providecommand \doibase [0]{https://doi.org/}%
\providecommand \selectlanguage [0]{\@gobble}%
\providecommand \bibinfo  [0]{\@secondoftwo}%
\providecommand \bibfield  [0]{\@secondoftwo}%
\providecommand \translation [1]{[#1]}%
\providecommand \BibitemOpen [0]{}%
\providecommand \bibitemStop [0]{}%
\providecommand \bibitemNoStop [0]{.\EOS\space}%
\providecommand \EOS [0]{\spacefactor3000\relax}%
\providecommand \BibitemShut  [1]{\csname bibitem#1\endcsname}%
\let\auto@bib@innerbib\@empty
\bibitem [{eli()}]{eli}%
  \BibitemOpen
  \href@noop {} {\bibinfo {title} {Eli white book}},\ \bibinfo {howpublished}
  {https://eli-laser.eu/media/1019/eli-whitebook.pdf},\ \bibinfo {note}
  {accessed: 2021-08-10}\BibitemShut {NoStop}%
\bibitem [{xce()}]{xcels}%
  \BibitemOpen
  \href@noop {} {\bibinfo {title} {Xcels white book}},\ \bibinfo {howpublished}
  {https://xcels.ipfran.ru/img/XCELS-Project-english-version.pdf},\ \bibinfo
  {note} {accessed: 2021-08-10}\BibitemShut {NoStop}%
\bibitem [{\citenamefont {Kitagawa}\ \emph {et~al.}(2004)\citenamefont
  {Kitagawa}, \citenamefont {Fujita}, \citenamefont {Kodama}, \citenamefont
  {Yoshida}, \citenamefont {Matsuo}, \citenamefont {Jitsuno}, \citenamefont
  {Kawasaki}, \citenamefont {Kitamura}, \citenamefont {Kanabe}, \citenamefont
  {Sakabe}, \citenamefont {Shigemori}, \citenamefont {Miyanaga},\ and\
  \citenamefont {Izawa}}]{kitagawa.qe.2004}%
  \BibitemOpen
  \bibfield  {author} {\bibinfo {author} {\bibfnamefont {Y.}~\bibnamefont
  {Kitagawa}}, \bibinfo {author} {\bibfnamefont {H.}~\bibnamefont {Fujita}},
  \bibinfo {author} {\bibfnamefont {R.}~\bibnamefont {Kodama}}, \bibinfo
  {author} {\bibfnamefont {H.}~\bibnamefont {Yoshida}}, \bibinfo {author}
  {\bibfnamefont {S.}~\bibnamefont {Matsuo}}, \bibinfo {author} {\bibfnamefont
  {T.}~\bibnamefont {Jitsuno}}, \bibinfo {author} {\bibfnamefont
  {T.}~\bibnamefont {Kawasaki}}, \bibinfo {author} {\bibfnamefont
  {H.}~\bibnamefont {Kitamura}}, \bibinfo {author} {\bibfnamefont
  {T.}~\bibnamefont {Kanabe}}, \bibinfo {author} {\bibfnamefont
  {S.}~\bibnamefont {Sakabe}}, \bibinfo {author} {\bibfnamefont
  {K.}~\bibnamefont {Shigemori}}, \bibinfo {author} {\bibfnamefont
  {N.}~\bibnamefont {Miyanaga}},\ and\ \bibinfo {author} {\bibfnamefont
  {Y.}~\bibnamefont {Izawa}},\ }\bibfield  {title} {\bibinfo {title}
  {Prepulse-free petawatt laser for a fast ignitor},\ }\href
  {https://doi.org/10.1109/jqe.2003.823043} {\bibfield  {journal} {\bibinfo
  {journal} {{IEEE} Journal of Quantum Electronics}\ }\textbf {\bibinfo
  {volume} {40}},\ \bibinfo {pages} {281} (\bibinfo {year} {2004})}\BibitemShut
  {NoStop}%
\bibitem [{\citenamefont {Kawanaka}\ \emph {et~al.}(2016)\citenamefont
  {Kawanaka}, \citenamefont {Tsubakimoto}, \citenamefont {Yoshida},
  \citenamefont {Fujioka}, \citenamefont {Fujimoto}, \citenamefont {Tokita},
  \citenamefont {Jitsuno},\ and\ \citenamefont {and}}]{kawanaka.jpcs.2016}%
  \BibitemOpen
  \bibfield  {author} {\bibinfo {author} {\bibfnamefont {J.}~\bibnamefont
  {Kawanaka}}, \bibinfo {author} {\bibfnamefont {K.}~\bibnamefont
  {Tsubakimoto}}, \bibinfo {author} {\bibfnamefont {H.}~\bibnamefont
  {Yoshida}}, \bibinfo {author} {\bibfnamefont {K.}~\bibnamefont {Fujioka}},
  \bibinfo {author} {\bibfnamefont {Y.}~\bibnamefont {Fujimoto}}, \bibinfo
  {author} {\bibfnamefont {S.}~\bibnamefont {Tokita}}, \bibinfo {author}
  {\bibfnamefont {T.}~\bibnamefont {Jitsuno}},\ and\ \bibinfo {author}
  {\bibfnamefont {N.~M.}\ \bibnamefont {and}},\ }\bibfield  {title} {\bibinfo
  {title} {Conceptual design of sub-exa-watt system by using optical parametric
  chirped pulse amplification},\ }\href
  {https://doi.org/10.1088/1742-6596/688/1/012044} {\bibfield  {journal}
  {\bibinfo  {journal} {Journal of Physics: Conference Series}\ }\textbf
  {\bibinfo {volume} {688}},\ \bibinfo {pages} {012044} (\bibinfo {year}
  {2016})}\BibitemShut {NoStop}%
\bibitem [{\citenamefont {Danson}\ \emph {et~al.}(2019)\citenamefont {Danson},
  \citenamefont {Haefner}, \citenamefont {Bromage}, \citenamefont {Butcher},
  \citenamefont {Chanteloup}, \citenamefont {Chowdhury}, \citenamefont
  {Galvanauskas}, \citenamefont {Gizzi}, \citenamefont {Hein}, \citenamefont
  {Hillier}, \citenamefont {Hopps}, \citenamefont {Kato}, \citenamefont
  {Khazanov}, \citenamefont {Kodama}, \citenamefont {Korn}, \citenamefont {Li},
  \citenamefont {Li}, \citenamefont {Limpert}, \citenamefont {Ma},
  \citenamefont {Nam}, \citenamefont {Neely}, \citenamefont {Papadopoulos},
  \citenamefont {Penman}, \citenamefont {Qian}, \citenamefont {Rocca},
  \citenamefont {Shaykin}, \citenamefont {Siders}, \citenamefont {Spindloe},
  \citenamefont {Szatm{\'{a}}ri}, \citenamefont {Trines}, \citenamefont {Zhu},
  \citenamefont {Zhu},\ and\ \citenamefont {Zuegel}}]{danson.hplse.2019}%
  \BibitemOpen
  \bibfield  {author} {\bibinfo {author} {\bibfnamefont {C.~N.}\ \bibnamefont
  {Danson}}, \bibinfo {author} {\bibfnamefont {C.}~\bibnamefont {Haefner}},
  \bibinfo {author} {\bibfnamefont {J.}~\bibnamefont {Bromage}}, \bibinfo
  {author} {\bibfnamefont {T.}~\bibnamefont {Butcher}}, \bibinfo {author}
  {\bibfnamefont {J.-C.~F.}\ \bibnamefont {Chanteloup}}, \bibinfo {author}
  {\bibfnamefont {E.~A.}\ \bibnamefont {Chowdhury}}, \bibinfo {author}
  {\bibfnamefont {A.}~\bibnamefont {Galvanauskas}}, \bibinfo {author}
  {\bibfnamefont {L.~A.}\ \bibnamefont {Gizzi}}, \bibinfo {author}
  {\bibfnamefont {J.}~\bibnamefont {Hein}}, \bibinfo {author} {\bibfnamefont
  {D.~I.}\ \bibnamefont {Hillier}}, \bibinfo {author} {\bibfnamefont {N.~W.}\
  \bibnamefont {Hopps}}, \bibinfo {author} {\bibfnamefont {Y.}~\bibnamefont
  {Kato}}, \bibinfo {author} {\bibfnamefont {E.~A.}\ \bibnamefont {Khazanov}},
  \bibinfo {author} {\bibfnamefont {R.}~\bibnamefont {Kodama}}, \bibinfo
  {author} {\bibfnamefont {G.}~\bibnamefont {Korn}}, \bibinfo {author}
  {\bibfnamefont {R.}~\bibnamefont {Li}}, \bibinfo {author} {\bibfnamefont
  {Y.}~\bibnamefont {Li}}, \bibinfo {author} {\bibfnamefont {J.}~\bibnamefont
  {Limpert}}, \bibinfo {author} {\bibfnamefont {J.}~\bibnamefont {Ma}},
  \bibinfo {author} {\bibfnamefont {C.~H.}\ \bibnamefont {Nam}}, \bibinfo
  {author} {\bibfnamefont {D.}~\bibnamefont {Neely}}, \bibinfo {author}
  {\bibfnamefont {D.}~\bibnamefont {Papadopoulos}}, \bibinfo {author}
  {\bibfnamefont {R.~R.}\ \bibnamefont {Penman}}, \bibinfo {author}
  {\bibfnamefont {L.}~\bibnamefont {Qian}}, \bibinfo {author} {\bibfnamefont
  {J.~J.}\ \bibnamefont {Rocca}}, \bibinfo {author} {\bibfnamefont {A.~A.}\
  \bibnamefont {Shaykin}}, \bibinfo {author} {\bibfnamefont {C.~W.}\
  \bibnamefont {Siders}}, \bibinfo {author} {\bibfnamefont {C.}~\bibnamefont
  {Spindloe}}, \bibinfo {author} {\bibfnamefont {S.}~\bibnamefont
  {Szatm{\'{a}}ri}}, \bibinfo {author} {\bibfnamefont {R.~M. G.~M.}\
  \bibnamefont {Trines}}, \bibinfo {author} {\bibfnamefont {J.}~\bibnamefont
  {Zhu}}, \bibinfo {author} {\bibfnamefont {P.}~\bibnamefont {Zhu}},\ and\
  \bibinfo {author} {\bibfnamefont {J.~D.}\ \bibnamefont {Zuegel}},\ }\bibfield
   {title} {\bibinfo {title} {Petawatt and exawatt class lasers worldwide},\
  }\bibfield  {journal} {\bibinfo  {journal} {High Power Laser Science and
  Engineering}\ }\textbf {\bibinfo {volume} {7}},\ \href
  {https://doi.org/10.1017/hpl.2019.36} {10.1017/hpl.2019.36} (\bibinfo {year}
  {2019})\BibitemShut {NoStop}%
\bibitem [{\citenamefont {Di~Piazza}\ \emph {et~al.}(2012)\citenamefont
  {Di~Piazza}, \citenamefont {Muller}, \citenamefont {Hatsagortsyan},\ and\
  \citenamefont {Keitel}}]{dipiazza.rmp.2012}%
  \BibitemOpen
  \bibfield  {author} {\bibinfo {author} {\bibfnamefont {A.}~\bibnamefont
  {Di~Piazza}}, \bibinfo {author} {\bibfnamefont {C.}~\bibnamefont {Muller}},
  \bibinfo {author} {\bibfnamefont {K.~Z.}\ \bibnamefont {Hatsagortsyan}},\
  and\ \bibinfo {author} {\bibfnamefont {C.~H.}\ \bibnamefont {Keitel}},\
  }\bibfield  {title} {\bibinfo {title} {{Extremely high-intensity laser
  interactions with fundamental quantum systems}},\ }\href
  {https://doi.org/10.1103/RevModPhys.84.1177} {\bibfield  {journal} {\bibinfo
  {journal} {Rev. Mod. Phys.}\ }\textbf {\bibinfo {volume} {84}},\ \bibinfo
  {pages} {1177} (\bibinfo {year} {2012})}\BibitemShut {NoStop}%
\bibitem [{\citenamefont {Gonoskov}\ \emph {et~al.}(2022)\citenamefont
  {Gonoskov}, \citenamefont {Blackburn}, \citenamefont {Marklund},\ and\
  \citenamefont {Bulanov}}]{gonoskov.rmp.2022}%
  \BibitemOpen
  \bibfield  {author} {\bibinfo {author} {\bibfnamefont {A.}~\bibnamefont
  {Gonoskov}}, \bibinfo {author} {\bibfnamefont {T.}~\bibnamefont {Blackburn}},
  \bibinfo {author} {\bibfnamefont {M.}~\bibnamefont {Marklund}},\ and\
  \bibinfo {author} {\bibfnamefont {S.}~\bibnamefont {Bulanov}},\ }\bibfield
  {title} {\bibinfo {title} {Charged particle motion and radiation in strong
  electromagnetic fields},\ }\bibfield  {journal} {\bibinfo  {journal} {Reviews
  of Modern Physics}\ }\textbf {\bibinfo {volume} {94}},\ \href
  {https://doi.org/10.1103/revmodphys.94.045001} {10.1103/revmodphys.94.045001}
  (\bibinfo {year} {2022})\BibitemShut {NoStop}%
\bibitem [{\citenamefont {Fedotov}\ \emph {et~al.}(2022)\citenamefont
  {Fedotov}, \citenamefont {Ilderton}, \citenamefont {Karbstein}, \citenamefont
  {King}, \citenamefont {Seipt}, \citenamefont {Taya},\ and\ \citenamefont
  {Torgrimsson}}]{fedotov.arxiv.2022}%
  \BibitemOpen
  \bibfield  {author} {\bibinfo {author} {\bibfnamefont {A.}~\bibnamefont
  {Fedotov}}, \bibinfo {author} {\bibfnamefont {A.}~\bibnamefont {Ilderton}},
  \bibinfo {author} {\bibfnamefont {F.}~\bibnamefont {Karbstein}}, \bibinfo
  {author} {\bibfnamefont {B.}~\bibnamefont {King}}, \bibinfo {author}
  {\bibfnamefont {D.}~\bibnamefont {Seipt}}, \bibinfo {author} {\bibfnamefont
  {H.}~\bibnamefont {Taya}},\ and\ \bibinfo {author} {\bibfnamefont
  {G.}~\bibnamefont {Torgrimsson}},\ }\href@noop {} {\bibinfo {title} {Advances
  in qed with intense background fields}} (\bibinfo {year} {2022}),\ \Eprint
  {https://arxiv.org/abs/arXiv:2203.00019} {arXiv:2203.00019} \BibitemShut
  {NoStop}%
\bibitem [{\citenamefont {Ritus}(1972)}]{ritus.ap.1972}%
  \BibitemOpen
  \bibfield  {author} {\bibinfo {author} {\bibfnamefont {V.}~\bibnamefont
  {Ritus}},\ }\bibfield  {title} {\bibinfo {title} {Radiative corrections in
  quantum electrodynamics with intense field and their analytical properties},\
  }\href {https://doi.org/10.1016/0003-4916(72)90191-1} {\bibfield  {journal}
  {\bibinfo  {journal} {Annals of Physics}\ }\textbf {\bibinfo {volume} {69}},\
  \bibinfo {pages} {555} (\bibinfo {year} {1972})}\BibitemShut {NoStop}%
\bibitem [{\citenamefont {Narozhny}(1980)}]{narozhny.prd.1980}%
  \BibitemOpen
  \bibfield  {author} {\bibinfo {author} {\bibfnamefont {N.~B.}\ \bibnamefont
  {Narozhny}},\ }\bibfield  {title} {\bibinfo {title} {Expansion parameter of
  perturbation theory in intense-field quantum electrodynamics},\ }\href
  {https://doi.org/10.1103/physrevd.21.1176} {\bibfield  {journal} {\bibinfo
  {journal} {Physical Review D}\ }\textbf {\bibinfo {volume} {21}},\ \bibinfo
  {pages} {1176} (\bibinfo {year} {1980})}\BibitemShut {NoStop}%
\bibitem [{\citenamefont {Fedotov}(2017)}]{fedotov.jpcs.2017}%
  \BibitemOpen
  \bibfield  {author} {\bibinfo {author} {\bibfnamefont {A.}~\bibnamefont
  {Fedotov}},\ }\bibfield  {title} {\bibinfo {title} {Conjecture of
  perturbative {QED} breakdown at$\alpha\chi^{2/3}\gtrsim$ 1},\ }\href
  {https://doi.org/10.1088/1742-6596/826/1/012027} {\bibfield  {journal}
  {\bibinfo  {journal} {Journal of Physics: Conference Series}\ }\textbf
  {\bibinfo {volume} {826}},\ \bibinfo {pages} {012027} (\bibinfo {year}
  {2017})}\BibitemShut {NoStop}%
\bibitem [{\citenamefont {Bula}\ \emph {et~al.}(1996)\citenamefont {Bula},
  \citenamefont {McDonald}, \citenamefont {Prebys}, \citenamefont {Bamber},
  \citenamefont {Boege}, \citenamefont {Kotseroglou}, \citenamefont
  {Melissinos}, \citenamefont {Meyerhofer}, \citenamefont {Ragg}, \citenamefont
  {Burke}, \citenamefont {Field}, \citenamefont {Horton-Smith}, \citenamefont
  {Odian}, \citenamefont {Spencer}, \citenamefont {Walz}, \citenamefont
  {Berridge}, \citenamefont {Bugg}, \citenamefont {Shmakov},\ and\
  \citenamefont {Weidemann}}]{bula.prl.1996}%
  \BibitemOpen
  \bibfield  {author} {\bibinfo {author} {\bibfnamefont {C.}~\bibnamefont
  {Bula}}, \bibinfo {author} {\bibfnamefont {K.~T.}\ \bibnamefont {McDonald}},
  \bibinfo {author} {\bibfnamefont {E.~J.}\ \bibnamefont {Prebys}}, \bibinfo
  {author} {\bibfnamefont {C.}~\bibnamefont {Bamber}}, \bibinfo {author}
  {\bibfnamefont {S.}~\bibnamefont {Boege}}, \bibinfo {author} {\bibfnamefont
  {T.}~\bibnamefont {Kotseroglou}}, \bibinfo {author} {\bibfnamefont {A.~C.}\
  \bibnamefont {Melissinos}}, \bibinfo {author} {\bibfnamefont {D.~D.}\
  \bibnamefont {Meyerhofer}}, \bibinfo {author} {\bibfnamefont
  {W.}~\bibnamefont {Ragg}}, \bibinfo {author} {\bibfnamefont {D.~L.}\
  \bibnamefont {Burke}}, \bibinfo {author} {\bibfnamefont {R.~C.}\ \bibnamefont
  {Field}}, \bibinfo {author} {\bibfnamefont {G.}~\bibnamefont {Horton-Smith}},
  \bibinfo {author} {\bibfnamefont {A.~C.}\ \bibnamefont {Odian}}, \bibinfo
  {author} {\bibfnamefont {J.~E.}\ \bibnamefont {Spencer}}, \bibinfo {author}
  {\bibfnamefont {D.}~\bibnamefont {Walz}}, \bibinfo {author} {\bibfnamefont
  {S.~C.}\ \bibnamefont {Berridge}}, \bibinfo {author} {\bibfnamefont {W.~M.}\
  \bibnamefont {Bugg}}, \bibinfo {author} {\bibfnamefont {K.}~\bibnamefont
  {Shmakov}},\ and\ \bibinfo {author} {\bibfnamefont {A.~W.}\ \bibnamefont
  {Weidemann}},\ }\bibfield  {title} {\bibinfo {title} {{Observation of
  nonlinear effects in compton scattering}},\ }\href
  {https://doi.org/10.1103/PhysRevLett.76.3116} {\bibfield  {journal} {\bibinfo
   {journal} {Physical Review Letters}\ }\textbf {\bibinfo {volume} {76}},\
  \bibinfo {pages} {3116} (\bibinfo {year} {1996})}\BibitemShut {NoStop}%
\bibitem [{\citenamefont {Burke}\ \emph {et~al.}(1997)\citenamefont {Burke},
  \citenamefont {Field}, \citenamefont {Horton-Smith}, \citenamefont {Spencer},
  \citenamefont {Walz}, \citenamefont {Berridge}, \citenamefont {Bugg},
  \citenamefont {Shmakov}, \citenamefont {Weidemann}, \citenamefont {Bula},
  \citenamefont {Mc~Donald}, \citenamefont {Prebys}, \citenamefont {Bamber},
  \citenamefont {Boege}, \citenamefont {Koffas}, \citenamefont {Kotseroglou},
  \citenamefont {Melissinos}, \citenamefont {Meyerhofer}, \citenamefont
  {Reis},\ and\ \citenamefont {Ragg}}]{burke.prl.1997}%
  \BibitemOpen
  \bibfield  {author} {\bibinfo {author} {\bibfnamefont {D.~L.}\ \bibnamefont
  {Burke}}, \bibinfo {author} {\bibfnamefont {R.~C.}\ \bibnamefont {Field}},
  \bibinfo {author} {\bibfnamefont {G.}~\bibnamefont {Horton-Smith}}, \bibinfo
  {author} {\bibfnamefont {J.~E.}\ \bibnamefont {Spencer}}, \bibinfo {author}
  {\bibfnamefont {D.}~\bibnamefont {Walz}}, \bibinfo {author} {\bibfnamefont
  {S.~C.}\ \bibnamefont {Berridge}}, \bibinfo {author} {\bibfnamefont {W.~M.}\
  \bibnamefont {Bugg}}, \bibinfo {author} {\bibfnamefont {K.}~\bibnamefont
  {Shmakov}}, \bibinfo {author} {\bibfnamefont {A.~W.}\ \bibnamefont
  {Weidemann}}, \bibinfo {author} {\bibfnamefont {C.}~\bibnamefont {Bula}},
  \bibinfo {author} {\bibfnamefont {K.~T.}\ \bibnamefont {Mc~Donald}}, \bibinfo
  {author} {\bibfnamefont {E.~J.}\ \bibnamefont {Prebys}}, \bibinfo {author}
  {\bibfnamefont {C.}~\bibnamefont {Bamber}}, \bibinfo {author} {\bibfnamefont
  {S.~J.}\ \bibnamefont {Boege}}, \bibinfo {author} {\bibfnamefont
  {T.}~\bibnamefont {Koffas}}, \bibinfo {author} {\bibfnamefont
  {T.}~\bibnamefont {Kotseroglou}}, \bibinfo {author} {\bibfnamefont {A.~C.}\
  \bibnamefont {Melissinos}}, \bibinfo {author} {\bibfnamefont {D.~D.}\
  \bibnamefont {Meyerhofer}}, \bibinfo {author} {\bibfnamefont {D.~A.}\
  \bibnamefont {Reis}},\ and\ \bibinfo {author} {\bibfnamefont
  {W.}~\bibnamefont {Ragg}},\ }\bibfield  {title} {\bibinfo {title} {{Positron
  production in multiphoton light-by-light scattering}},\ }\href
  {https://doi.org/10.1103/PhysRevLett.79.1626} {\bibfield  {journal} {\bibinfo
   {journal} {Physical Review Letters}\ }\textbf {\bibinfo {volume} {79}},\
  \bibinfo {pages} {1626} (\bibinfo {year} {1997})}\BibitemShut {NoStop}%
\bibitem [{\citenamefont {Yakimenko}\ \emph
  {et~al.}(2019{\natexlab{a}})\citenamefont {Yakimenko}, \citenamefont
  {Alsberg}, \citenamefont {Bong}, \citenamefont {Bouchard}, \citenamefont
  {Clarke}, \citenamefont {Emma}, \citenamefont {Green}, \citenamefont {Hast},
  \citenamefont {Hogan}, \citenamefont {Seabury}, \citenamefont {Lipkowitz},
  \citenamefont {O'Shea}, \citenamefont {Storey}, \citenamefont {White},\ and\
  \citenamefont {Yocky}}]{yakimenko.prab.2019}%
  \BibitemOpen
  \bibfield  {author} {\bibinfo {author} {\bibfnamefont {V.}~\bibnamefont
  {Yakimenko}}, \bibinfo {author} {\bibfnamefont {L.}~\bibnamefont {Alsberg}},
  \bibinfo {author} {\bibfnamefont {E.}~\bibnamefont {Bong}}, \bibinfo {author}
  {\bibfnamefont {G.}~\bibnamefont {Bouchard}}, \bibinfo {author}
  {\bibfnamefont {C.}~\bibnamefont {Clarke}}, \bibinfo {author} {\bibfnamefont
  {C.}~\bibnamefont {Emma}}, \bibinfo {author} {\bibfnamefont {S.}~\bibnamefont
  {Green}}, \bibinfo {author} {\bibfnamefont {C.}~\bibnamefont {Hast}},
  \bibinfo {author} {\bibfnamefont {M.~J.}\ \bibnamefont {Hogan}}, \bibinfo
  {author} {\bibfnamefont {J.}~\bibnamefont {Seabury}}, \bibinfo {author}
  {\bibfnamefont {N.}~\bibnamefont {Lipkowitz}}, \bibinfo {author}
  {\bibfnamefont {B.}~\bibnamefont {O'Shea}}, \bibinfo {author} {\bibfnamefont
  {D.}~\bibnamefont {Storey}}, \bibinfo {author} {\bibfnamefont
  {G.}~\bibnamefont {White}},\ and\ \bibinfo {author} {\bibfnamefont
  {G.}~\bibnamefont {Yocky}},\ }\bibfield  {title} {\bibinfo {title} {{FACET-II
  facility for advanced accelerator experimental tests}},\ }\href
  {https://doi.org/10.1103/PhysRevAccelBeams.22.101301} {\bibfield  {journal}
  {\bibinfo  {journal} {Physical Review Accelerators and Beams}\ }\textbf
  {\bibinfo {volume} {22}},\ \bibinfo {pages} {101301} (\bibinfo {year}
  {2019}{\natexlab{a}})}\BibitemShut {NoStop}%
\bibitem [{\citenamefont {Abramowicz}\ \emph {et~al.}(2019)\citenamefont
  {Abramowicz}, \citenamefont {Altarelli}, \citenamefont {A{\ss}mann},
  \citenamefont {Behnke}, \citenamefont {Benhammou}, \citenamefont {Borysov},
  \citenamefont {Borysova}, \citenamefont {Brinkmann}, \citenamefont {Burkart},
  \citenamefont {B{\"{u}}{\ss}er}, \citenamefont {Davidi}, \citenamefont
  {Decking}, \citenamefont {Elkina}, \citenamefont {Harsh}, \citenamefont
  {Hartin}, \citenamefont {Hartl}, \citenamefont {Heinemann}, \citenamefont
  {Heinzl}, \citenamefont {Tal~Hod}, \citenamefont {Hoffmann}, \citenamefont
  {Ilderton}, \citenamefont {King}, \citenamefont {Levy}, \citenamefont {List},
  \citenamefont {Maier}, \citenamefont {Negodin}, \citenamefont {Perez},
  \citenamefont {Pomerantz}, \citenamefont {Ringwald}, \citenamefont
  {R{\"{o}}del}, \citenamefont {Saimpert}, \citenamefont {Salgado},
  \citenamefont {Sarri}, \citenamefont {Savoray}, \citenamefont {Teter},
  \citenamefont {Wing},\ and\ \citenamefont {Zepf}}]{abramowicz.arxiv.2019}%
  \BibitemOpen
  \bibfield  {author} {\bibinfo {author} {\bibfnamefont {H.}~\bibnamefont
  {Abramowicz}}, \bibinfo {author} {\bibfnamefont {M.}~\bibnamefont
  {Altarelli}}, \bibinfo {author} {\bibfnamefont {R.}~\bibnamefont
  {A{\ss}mann}}, \bibinfo {author} {\bibfnamefont {T.}~\bibnamefont {Behnke}},
  \bibinfo {author} {\bibfnamefont {Y.}~\bibnamefont {Benhammou}}, \bibinfo
  {author} {\bibfnamefont {O.}~\bibnamefont {Borysov}}, \bibinfo {author}
  {\bibfnamefont {M.}~\bibnamefont {Borysova}}, \bibinfo {author}
  {\bibfnamefont {R.}~\bibnamefont {Brinkmann}}, \bibinfo {author}
  {\bibfnamefont {F.}~\bibnamefont {Burkart}}, \bibinfo {author} {\bibfnamefont
  {K.}~\bibnamefont {B{\"{u}}{\ss}er}}, \bibinfo {author} {\bibfnamefont
  {O.}~\bibnamefont {Davidi}}, \bibinfo {author} {\bibfnamefont
  {W.}~\bibnamefont {Decking}}, \bibinfo {author} {\bibfnamefont
  {N.}~\bibnamefont {Elkina}}, \bibinfo {author} {\bibfnamefont
  {H.}~\bibnamefont {Harsh}}, \bibinfo {author} {\bibfnamefont
  {A.}~\bibnamefont {Hartin}}, \bibinfo {author} {\bibfnamefont
  {I.}~\bibnamefont {Hartl}}, \bibinfo {author} {\bibfnamefont
  {B.}~\bibnamefont {Heinemann}}, \bibinfo {author} {\bibfnamefont
  {T.}~\bibnamefont {Heinzl}}, \bibinfo {author} {\bibfnamefont
  {N.}~\bibnamefont {Tal~Hod}}, \bibinfo {author} {\bibfnamefont
  {M.}~\bibnamefont {Hoffmann}}, \bibinfo {author} {\bibfnamefont
  {A.}~\bibnamefont {Ilderton}}, \bibinfo {author} {\bibfnamefont
  {B.}~\bibnamefont {King}}, \bibinfo {author} {\bibfnamefont {A.}~\bibnamefont
  {Levy}}, \bibinfo {author} {\bibfnamefont {J.}~\bibnamefont {List}}, \bibinfo
  {author} {\bibfnamefont {A.~R.}\ \bibnamefont {Maier}}, \bibinfo {author}
  {\bibfnamefont {E.}~\bibnamefont {Negodin}}, \bibinfo {author} {\bibfnamefont
  {G.}~\bibnamefont {Perez}}, \bibinfo {author} {\bibfnamefont
  {I.}~\bibnamefont {Pomerantz}}, \bibinfo {author} {\bibfnamefont
  {A.}~\bibnamefont {Ringwald}}, \bibinfo {author} {\bibfnamefont
  {C.}~\bibnamefont {R{\"{o}}del}}, \bibinfo {author} {\bibfnamefont
  {M.}~\bibnamefont {Saimpert}}, \bibinfo {author} {\bibfnamefont
  {F.}~\bibnamefont {Salgado}}, \bibinfo {author} {\bibfnamefont
  {G.}~\bibnamefont {Sarri}}, \bibinfo {author} {\bibfnamefont
  {I.}~\bibnamefont {Savoray}}, \bibinfo {author} {\bibfnamefont
  {T.}~\bibnamefont {Teter}}, \bibinfo {author} {\bibfnamefont
  {M.}~\bibnamefont {Wing}},\ and\ \bibinfo {author} {\bibfnamefont
  {M.}~\bibnamefont {Zepf}},\ }\href {http://arxiv.org/abs/1909.00860}
  {\bibinfo {title} {{Letter of intent for the LUXE experiment}}} (\bibinfo
  {year} {2019})\BibitemShut {NoStop}%
\bibitem [{\citenamefont {Cole}\ \emph {et~al.}(2018)\citenamefont {Cole},
  \citenamefont {Behm}, \citenamefont {Gerstmayr}, \citenamefont {Blackburn},
  \citenamefont {Wood}, \citenamefont {Baird}, \citenamefont {Duff},
  \citenamefont {Harvey}, \citenamefont {Ilderton}, \citenamefont {Joglekar},
  \citenamefont {Krushelnick}, \citenamefont {Kuschel}, \citenamefont
  {Marklund}, \citenamefont {McKenna}, \citenamefont {Murphy}, \citenamefont
  {Poder}, \citenamefont {Ridgers}, \citenamefont {Samarin}, \citenamefont
  {Sarri}, \citenamefont {Symes}, \citenamefont {Thomas}, \citenamefont
  {Warwick}, \citenamefont {Zepf}, \citenamefont {Najmudin},\ and\
  \citenamefont {Mangles}}]{cole.prx.2018}%
  \BibitemOpen
  \bibfield  {author} {\bibinfo {author} {\bibfnamefont {J.~M.}\ \bibnamefont
  {Cole}}, \bibinfo {author} {\bibfnamefont {K.~T.}\ \bibnamefont {Behm}},
  \bibinfo {author} {\bibfnamefont {E.}~\bibnamefont {Gerstmayr}}, \bibinfo
  {author} {\bibfnamefont {T.~G.}\ \bibnamefont {Blackburn}}, \bibinfo {author}
  {\bibfnamefont {J.~C.}\ \bibnamefont {Wood}}, \bibinfo {author}
  {\bibfnamefont {C.~D.}\ \bibnamefont {Baird}}, \bibinfo {author}
  {\bibfnamefont {M.~J.}\ \bibnamefont {Duff}}, \bibinfo {author}
  {\bibfnamefont {C.}~\bibnamefont {Harvey}}, \bibinfo {author} {\bibfnamefont
  {A.}~\bibnamefont {Ilderton}}, \bibinfo {author} {\bibfnamefont {A.~S.}\
  \bibnamefont {Joglekar}}, \bibinfo {author} {\bibfnamefont {K.}~\bibnamefont
  {Krushelnick}}, \bibinfo {author} {\bibfnamefont {S.}~\bibnamefont
  {Kuschel}}, \bibinfo {author} {\bibfnamefont {M.}~\bibnamefont {Marklund}},
  \bibinfo {author} {\bibfnamefont {P.}~\bibnamefont {McKenna}}, \bibinfo
  {author} {\bibfnamefont {C.~D.}\ \bibnamefont {Murphy}}, \bibinfo {author}
  {\bibfnamefont {K.}~\bibnamefont {Poder}}, \bibinfo {author} {\bibfnamefont
  {C.~P.}\ \bibnamefont {Ridgers}}, \bibinfo {author} {\bibfnamefont {G.~M.}\
  \bibnamefont {Samarin}}, \bibinfo {author} {\bibfnamefont {G.}~\bibnamefont
  {Sarri}}, \bibinfo {author} {\bibfnamefont {D.~R.}\ \bibnamefont {Symes}},
  \bibinfo {author} {\bibfnamefont {A.~G.}\ \bibnamefont {Thomas}}, \bibinfo
  {author} {\bibfnamefont {J.}~\bibnamefont {Warwick}}, \bibinfo {author}
  {\bibfnamefont {M.}~\bibnamefont {Zepf}}, \bibinfo {author} {\bibfnamefont
  {Z.}~\bibnamefont {Najmudin}},\ and\ \bibinfo {author} {\bibfnamefont
  {S.~P.}\ \bibnamefont {Mangles}},\ }\bibfield  {title} {\bibinfo {title}
  {{Experimental Evidence of Radiation Reaction in the Collision of a
  High-Intensity Laser Pulse with a Laser-Wakefield Accelerated Electron
  Beam}},\ }\bibfield  {journal} {\bibinfo  {journal} {Physical Review X}\
  }\textbf {\bibinfo {volume} {8}},\ \href
  {https://doi.org/10.1103/PhysRevX.8.011020} {10.1103/PhysRevX.8.011020}
  (\bibinfo {year} {2018})\BibitemShut {NoStop}%
\bibitem [{\citenamefont {Poder}\ \emph {et~al.}(2018)\citenamefont {Poder},
  \citenamefont {Tamburini}, \citenamefont {Sarri}, \citenamefont {Di~Piazza},
  \citenamefont {Kuschel}, \citenamefont {Baird}, \citenamefont {Behm},
  \citenamefont {Bohlen}, \citenamefont {Cole}, \citenamefont {Corvan},
  \citenamefont {Duff}, \citenamefont {Gerstmayr}, \citenamefont {Keitel},
  \citenamefont {Krushelnick}, \citenamefont {Mangles}, \citenamefont
  {McKenna}, \citenamefont {Murphy}, \citenamefont {Najmudin}, \citenamefont
  {Ridgers}, \citenamefont {Samarin}, \citenamefont {Symes}, \citenamefont
  {Thomas}, \citenamefont {Warwick},\ and\ \citenamefont
  {Zepf}}]{poder.prx.2018}%
  \BibitemOpen
  \bibfield  {author} {\bibinfo {author} {\bibfnamefont {K.}~\bibnamefont
  {Poder}}, \bibinfo {author} {\bibfnamefont {M.}~\bibnamefont {Tamburini}},
  \bibinfo {author} {\bibfnamefont {G.}~\bibnamefont {Sarri}}, \bibinfo
  {author} {\bibfnamefont {A.}~\bibnamefont {Di~Piazza}}, \bibinfo {author}
  {\bibfnamefont {S.}~\bibnamefont {Kuschel}}, \bibinfo {author} {\bibfnamefont
  {C.~D.}\ \bibnamefont {Baird}}, \bibinfo {author} {\bibfnamefont
  {K.}~\bibnamefont {Behm}}, \bibinfo {author} {\bibfnamefont {S.}~\bibnamefont
  {Bohlen}}, \bibinfo {author} {\bibfnamefont {J.~M.}\ \bibnamefont {Cole}},
  \bibinfo {author} {\bibfnamefont {D.~J.}\ \bibnamefont {Corvan}}, \bibinfo
  {author} {\bibfnamefont {M.}~\bibnamefont {Duff}}, \bibinfo {author}
  {\bibfnamefont {E.}~\bibnamefont {Gerstmayr}}, \bibinfo {author}
  {\bibfnamefont {C.~H.}\ \bibnamefont {Keitel}}, \bibinfo {author}
  {\bibfnamefont {K.}~\bibnamefont {Krushelnick}}, \bibinfo {author}
  {\bibfnamefont {S.~P.}\ \bibnamefont {Mangles}}, \bibinfo {author}
  {\bibfnamefont {P.}~\bibnamefont {McKenna}}, \bibinfo {author} {\bibfnamefont
  {C.~D.}\ \bibnamefont {Murphy}}, \bibinfo {author} {\bibfnamefont
  {Z.}~\bibnamefont {Najmudin}}, \bibinfo {author} {\bibfnamefont {C.~P.}\
  \bibnamefont {Ridgers}}, \bibinfo {author} {\bibfnamefont {G.~M.}\
  \bibnamefont {Samarin}}, \bibinfo {author} {\bibfnamefont {D.~R.}\
  \bibnamefont {Symes}}, \bibinfo {author} {\bibfnamefont {A.~G.}\ \bibnamefont
  {Thomas}}, \bibinfo {author} {\bibfnamefont {J.}~\bibnamefont {Warwick}},\
  and\ \bibinfo {author} {\bibfnamefont {M.}~\bibnamefont {Zepf}},\ }\bibfield
  {title} {\bibinfo {title} {{Experimental Signatures of the Quantum Nature of
  Radiation Reaction in the Field of an Ultraintense Laser}},\ }\href
  {https://doi.org/10.1103/PhysRevX.8.031004} {\bibfield  {journal} {\bibinfo
  {journal} {Physical Review X}\ }\textbf {\bibinfo {volume} {8}},\ \bibinfo
  {pages} {31004} (\bibinfo {year} {2018})}\BibitemShut {NoStop}%
\bibitem [{\citenamefont {Gonoskov}\ \emph {et~al.}(2014)\citenamefont
  {Gonoskov}, \citenamefont {Bashinov}, \citenamefont {Gonoskov}, \citenamefont
  {Harvey}, \citenamefont {Ilderton}, \citenamefont {Kim}, \citenamefont
  {Marklund}, \citenamefont {Mourou},\ and\ \citenamefont
  {Sergeev}}]{gonoskov.prl.2014}%
  \BibitemOpen
  \bibfield  {author} {\bibinfo {author} {\bibfnamefont {A.}~\bibnamefont
  {Gonoskov}}, \bibinfo {author} {\bibfnamefont {A.}~\bibnamefont {Bashinov}},
  \bibinfo {author} {\bibfnamefont {I.}~\bibnamefont {Gonoskov}}, \bibinfo
  {author} {\bibfnamefont {C.}~\bibnamefont {Harvey}}, \bibinfo {author}
  {\bibfnamefont {A.}~\bibnamefont {Ilderton}}, \bibinfo {author}
  {\bibfnamefont {A.}~\bibnamefont {Kim}}, \bibinfo {author} {\bibfnamefont
  {M.}~\bibnamefont {Marklund}}, \bibinfo {author} {\bibfnamefont
  {G.}~\bibnamefont {Mourou}},\ and\ \bibinfo {author} {\bibfnamefont
  {A.}~\bibnamefont {Sergeev}},\ }\bibfield  {title} {\bibinfo {title}
  {{Anomalous Radiative Trapping in Laser Fields of Extreme Intensity}},\
  }\href {https://doi.org/10.1103/PhysRevLett.113.014801} {\bibfield  {journal}
  {\bibinfo  {journal} {Physical Review Letters}\ }\textbf {\bibinfo {volume}
  {113}},\ \bibinfo {pages} {014801} (\bibinfo {year} {2014})}\BibitemShut
  {NoStop}%
\bibitem [{\citenamefont {Bulanov}\ \emph
  {et~al.}(2010{\natexlab{a}})\citenamefont {Bulanov}, \citenamefont
  {Esirkepov}, \citenamefont {Thomas}, \citenamefont {Koga},\ and\
  \citenamefont {Bulanov}}]{bulanov.prl.2010b}%
  \BibitemOpen
  \bibfield  {author} {\bibinfo {author} {\bibfnamefont {S.~S.}\ \bibnamefont
  {Bulanov}}, \bibinfo {author} {\bibfnamefont {T.~Z.}\ \bibnamefont
  {Esirkepov}}, \bibinfo {author} {\bibfnamefont {A.~G.~R.}\ \bibnamefont
  {Thomas}}, \bibinfo {author} {\bibfnamefont {J.~K.}\ \bibnamefont {Koga}},\
  and\ \bibinfo {author} {\bibfnamefont {S.~V.}\ \bibnamefont {Bulanov}},\
  }\bibfield  {title} {\bibinfo {title} {Schwinger limit attainability with
  extreme power lasers},\ }\bibfield  {journal} {\bibinfo  {journal} {Physical
  Review Letters}\ }\textbf {\bibinfo {volume} {105}},\ \href
  {https://doi.org/10.1103/physrevlett.105.220407}
  {10.1103/physrevlett.105.220407} (\bibinfo {year}
  {2010}{\natexlab{a}})\BibitemShut {NoStop}%
\bibitem [{\citenamefont {Yakimenko}\ \emph
  {et~al.}(2019{\natexlab{b}})\citenamefont {Yakimenko}, \citenamefont
  {Alsberg}, \citenamefont {Bong}, \citenamefont {Bouchard}, \citenamefont
  {Clarke}, \citenamefont {Emma}, \citenamefont {Green}, \citenamefont {Hast},
  \citenamefont {Hogan}, \citenamefont {Seabury}, \citenamefont {Lipkowitz},
  \citenamefont {O'Shea}, \citenamefont {Storey}, \citenamefont {White},\ and\
  \citenamefont {Yocky}}]{yakimenko.prl.2019}%
  \BibitemOpen
  \bibfield  {author} {\bibinfo {author} {\bibfnamefont {V.}~\bibnamefont
  {Yakimenko}}, \bibinfo {author} {\bibfnamefont {L.}~\bibnamefont {Alsberg}},
  \bibinfo {author} {\bibfnamefont {E.}~\bibnamefont {Bong}}, \bibinfo {author}
  {\bibfnamefont {G.}~\bibnamefont {Bouchard}}, \bibinfo {author}
  {\bibfnamefont {C.}~\bibnamefont {Clarke}}, \bibinfo {author} {\bibfnamefont
  {C.}~\bibnamefont {Emma}}, \bibinfo {author} {\bibfnamefont {S.}~\bibnamefont
  {Green}}, \bibinfo {author} {\bibfnamefont {C.}~\bibnamefont {Hast}},
  \bibinfo {author} {\bibfnamefont {M.}~\bibnamefont {Hogan}}, \bibinfo
  {author} {\bibfnamefont {J.}~\bibnamefont {Seabury}}, \bibinfo {author}
  {\bibfnamefont {N.}~\bibnamefont {Lipkowitz}}, \bibinfo {author}
  {\bibfnamefont {B.}~\bibnamefont {O'Shea}}, \bibinfo {author} {\bibfnamefont
  {D.}~\bibnamefont {Storey}}, \bibinfo {author} {\bibfnamefont
  {G.}~\bibnamefont {White}},\ and\ \bibinfo {author} {\bibfnamefont
  {G.}~\bibnamefont {Yocky}},\ }\bibfield  {title} {\bibinfo {title}
  {{FACET}-{II} facility for advanced accelerator experimental tests},\
  }\bibfield  {journal} {\bibinfo  {journal} {Physical Review Accelerators and
  Beams}\ }\textbf {\bibinfo {volume} {22}},\ \href
  {https://doi.org/10.1103/physrevaccelbeams.22.101301}
  {10.1103/physrevaccelbeams.22.101301} (\bibinfo {year}
  {2019}{\natexlab{b}})\BibitemShut {NoStop}%
\bibitem [{\citenamefont {Bassett}(1986)}]{bassett.oa.1986}%
  \BibitemOpen
  \bibfield  {author} {\bibinfo {author} {\bibfnamefont {I.~M.}\ \bibnamefont
  {Bassett}},\ }\bibfield  {title} {\bibinfo {title} {{Limit to Concentration
  by Focusing}},\ }\href {https://doi.org/10.1080/713821943} {\bibfield
  {journal} {\bibinfo  {journal} {Opt. Acta}\ }\textbf {\bibinfo {volume}
  {33}},\ \bibinfo {pages} {279} (\bibinfo {year} {1986})}\BibitemShut
  {NoStop}%
\bibitem [{\citenamefont {Gonoskov}\ \emph {et~al.}(2012)\citenamefont
  {Gonoskov}, \citenamefont {Aiello}, \citenamefont {Heugel},\ and\
  \citenamefont {Leuchs}}]{gonoskov.pra.2012}%
  \BibitemOpen
  \bibfield  {author} {\bibinfo {author} {\bibfnamefont {I.}~\bibnamefont
  {Gonoskov}}, \bibinfo {author} {\bibfnamefont {A.}~\bibnamefont {Aiello}},
  \bibinfo {author} {\bibfnamefont {S.}~\bibnamefont {Heugel}},\ and\ \bibinfo
  {author} {\bibfnamefont {G.}~\bibnamefont {Leuchs}},\ }\bibfield  {title}
  {\bibinfo {title} {{Dipole pulse theory: Maximizing the field amplitude from
  4{$\pi$} focused laser pulses}},\ }\href
  {https://doi.org/10.1103/PhysRevA.86.053836} {\bibfield  {journal} {\bibinfo
  {journal} {Physical Review A - Atomic, Molecular, and Optical Physics}\
  }\textbf {\bibinfo {volume} {86}},\ \bibinfo {pages} {053836} (\bibinfo
  {year} {2012})}\BibitemShut {NoStop}%
\bibitem [{\citenamefont {Jeong}\ \emph {et~al.}(2020)\citenamefont {Jeong},
  \citenamefont {Bulanov}, \citenamefont {Sasorov}, \citenamefont {Bulanov},
  \citenamefont {Koga},\ and\ \citenamefont {Korn}}]{jeong.oe.2020}%
  \BibitemOpen
  \bibfield  {author} {\bibinfo {author} {\bibfnamefont {T.~M.}\ \bibnamefont
  {Jeong}}, \bibinfo {author} {\bibfnamefont {S.~V.}\ \bibnamefont {Bulanov}},
  \bibinfo {author} {\bibfnamefont {P.~V.}\ \bibnamefont {Sasorov}}, \bibinfo
  {author} {\bibfnamefont {S.~S.}\ \bibnamefont {Bulanov}}, \bibinfo {author}
  {\bibfnamefont {J.~K.}\ \bibnamefont {Koga}},\ and\ \bibinfo {author}
  {\bibfnamefont {G.}~\bibnamefont {Korn}},\ }\bibfield  {title} {\bibinfo
  {title} {4$\uppi$-spherically focused electromagnetic wave: diffraction
  optics approach and high-power limits},\ }\href
  {https://doi.org/10.1364/oe.387654} {\bibfield  {journal} {\bibinfo
  {journal} {Optics Express}\ }\textbf {\bibinfo {volume} {28}},\ \bibinfo
  {pages} {13991} (\bibinfo {year} {2020})}\BibitemShut {NoStop}%
\bibitem [{\citenamefont {Bulanov}\ \emph
  {et~al.}(2010{\natexlab{b}})\citenamefont {Bulanov}, \citenamefont {Mur},
  \citenamefont {Narozhny}, \citenamefont {Nees},\ and\ \citenamefont
  {Popov}}]{bulanov.prl.2010}%
  \BibitemOpen
  \bibfield  {author} {\bibinfo {author} {\bibfnamefont {S.~S.}\ \bibnamefont
  {Bulanov}}, \bibinfo {author} {\bibfnamefont {V.~D.}\ \bibnamefont {Mur}},
  \bibinfo {author} {\bibfnamefont {N.~B.}\ \bibnamefont {Narozhny}}, \bibinfo
  {author} {\bibfnamefont {J.}~\bibnamefont {Nees}},\ and\ \bibinfo {author}
  {\bibfnamefont {V.~S.}\ \bibnamefont {Popov}},\ }\bibfield  {title} {\bibinfo
  {title} {Multiple colliding electromagnetic pulses: A way to lower the
  threshold of $e^+ e^-$ pair production from vacuum},\ }\bibfield  {journal}
  {\bibinfo  {journal} {Physical Review Letters}\ }\textbf {\bibinfo {volume}
  {104}},\ \href {https://doi.org/10.1103/physrevlett.104.220404}
  {10.1103/physrevlett.104.220404} (\bibinfo {year}
  {2010}{\natexlab{b}})\BibitemShut {NoStop}%
\bibitem [{\citenamefont {Gonoskov}\ \emph {et~al.}(2017)\citenamefont
  {Gonoskov}, \citenamefont {Bashinov}, \citenamefont {Bastrakov},
  \citenamefont {Efimenko}, \citenamefont {Ilderton}, \citenamefont {Kim},
  \citenamefont {Marklund}, \citenamefont {Meyerov}, \citenamefont {Muraviev},\
  and\ \citenamefont {Sergeev}}]{gonoskov.prx.2017}%
  \BibitemOpen
  \bibfield  {author} {\bibinfo {author} {\bibfnamefont {A.}~\bibnamefont
  {Gonoskov}}, \bibinfo {author} {\bibfnamefont {A.}~\bibnamefont {Bashinov}},
  \bibinfo {author} {\bibfnamefont {S.}~\bibnamefont {Bastrakov}}, \bibinfo
  {author} {\bibfnamefont {E.}~\bibnamefont {Efimenko}}, \bibinfo {author}
  {\bibfnamefont {A.}~\bibnamefont {Ilderton}}, \bibinfo {author}
  {\bibfnamefont {A.}~\bibnamefont {Kim}}, \bibinfo {author} {\bibfnamefont
  {M.}~\bibnamefont {Marklund}}, \bibinfo {author} {\bibfnamefont
  {I.}~\bibnamefont {Meyerov}}, \bibinfo {author} {\bibfnamefont
  {A.}~\bibnamefont {Muraviev}},\ and\ \bibinfo {author} {\bibfnamefont
  {A.}~\bibnamefont {Sergeev}},\ }\bibfield  {title} {\bibinfo {title}
  {{Ultrabright GeV photon source via controlled electromagnetic cascades in
  laser-dipole waves}},\ }\href {https://doi.org/10.1103/PhysRevX.7.041003}
  {\bibfield  {journal} {\bibinfo  {journal} {Physical Review X}\ }\textbf
  {\bibinfo {volume} {7}},\ \bibinfo {pages} {41003} (\bibinfo {year}
  {2017})}\BibitemShut {NoStop}%
\bibitem [{\citenamefont {Magnusson}\ \emph
  {et~al.}(2019{\natexlab{a}})\citenamefont {Magnusson}, \citenamefont
  {Gonoskov}, \citenamefont {Marklund}, \citenamefont {Esirkepov},
  \citenamefont {Koga}, \citenamefont {Kondo}, \citenamefont {Kando},
  \citenamefont {Bulanov}, \citenamefont {Korn}, \citenamefont {Geddes},
  \citenamefont {Schroeder}, \citenamefont {Esarey},\ and\ \citenamefont
  {Bulanov}}]{magnusson.pra.2019}%
  \BibitemOpen
  \bibfield  {author} {\bibinfo {author} {\bibfnamefont {J.}~\bibnamefont
  {Magnusson}}, \bibinfo {author} {\bibfnamefont {A.}~\bibnamefont {Gonoskov}},
  \bibinfo {author} {\bibfnamefont {M.}~\bibnamefont {Marklund}}, \bibinfo
  {author} {\bibfnamefont {T.~Z.}\ \bibnamefont {Esirkepov}}, \bibinfo {author}
  {\bibfnamefont {J.~K.}\ \bibnamefont {Koga}}, \bibinfo {author}
  {\bibfnamefont {K.}~\bibnamefont {Kondo}}, \bibinfo {author} {\bibfnamefont
  {M.}~\bibnamefont {Kando}}, \bibinfo {author} {\bibfnamefont {S.~V.}\
  \bibnamefont {Bulanov}}, \bibinfo {author} {\bibfnamefont {G.}~\bibnamefont
  {Korn}}, \bibinfo {author} {\bibfnamefont {C.~G.}\ \bibnamefont {Geddes}},
  \bibinfo {author} {\bibfnamefont {C.~B.}\ \bibnamefont {Schroeder}}, \bibinfo
  {author} {\bibfnamefont {E.}~\bibnamefont {Esarey}},\ and\ \bibinfo {author}
  {\bibfnamefont {S.~S.}\ \bibnamefont {Bulanov}},\ }\bibfield  {title}
  {\bibinfo {title} {{Multiple colliding laser pulses as a basis for studying
  high-field high-energy physics}},\ }\bibfield  {journal} {\bibinfo  {journal}
  {Physical Review A}\ }\textbf {\bibinfo {volume} {100}},\ \href
  {https://doi.org/10.1103/PhysRevA.100.063404} {10.1103/PhysRevA.100.063404}
  (\bibinfo {year} {2019}{\natexlab{a}})\BibitemShut {NoStop}%
\bibitem [{\citenamefont {Magnusson}\ \emph
  {et~al.}(2019{\natexlab{b}})\citenamefont {Magnusson}, \citenamefont
  {Gonoskov}, \citenamefont {Marklund}, \citenamefont {Esirkepov},
  \citenamefont {Koga}, \citenamefont {Kondo}, \citenamefont {Kando},
  \citenamefont {Bulanov}, \citenamefont {Korn},\ and\ \citenamefont
  {Bulanov}}]{magnusson.prl.2019}%
  \BibitemOpen
  \bibfield  {author} {\bibinfo {author} {\bibfnamefont {J.}~\bibnamefont
  {Magnusson}}, \bibinfo {author} {\bibfnamefont {A.}~\bibnamefont {Gonoskov}},
  \bibinfo {author} {\bibfnamefont {M.}~\bibnamefont {Marklund}}, \bibinfo
  {author} {\bibfnamefont {T.}~\bibnamefont {Esirkepov}}, \bibinfo {author}
  {\bibfnamefont {J.}~\bibnamefont {Koga}}, \bibinfo {author} {\bibfnamefont
  {K.}~\bibnamefont {Kondo}}, \bibinfo {author} {\bibfnamefont
  {M.}~\bibnamefont {Kando}}, \bibinfo {author} {\bibfnamefont
  {S.}~\bibnamefont {Bulanov}}, \bibinfo {author} {\bibfnamefont
  {G.}~\bibnamefont {Korn}},\ and\ \bibinfo {author} {\bibfnamefont
  {S.}~\bibnamefont {Bulanov}},\ }\bibfield  {title} {\bibinfo {title}
  {Laser-particle collider for multi-{GeV} photon production},\ }\bibfield
  {journal} {\bibinfo  {journal} {Physical Review Letters}\ }\textbf {\bibinfo
  {volume} {122}},\ \href {https://doi.org/10.1103/physrevlett.122.254801}
  {10.1103/physrevlett.122.254801} (\bibinfo {year}
  {2019}{\natexlab{b}})\BibitemShut {NoStop}%
\bibitem [{\citenamefont {Efimenko}\ \emph {et~al.}(2019)\citenamefont
  {Efimenko}, \citenamefont {Bashinov}, \citenamefont {Gonoskov}, \citenamefont
  {Bastrakov}, \citenamefont {Muraviev}, \citenamefont {Meyerov}, \citenamefont
  {Kim},\ and\ \citenamefont {Sergeev}}]{efimenko.pre.2019}%
  \BibitemOpen
  \bibfield  {author} {\bibinfo {author} {\bibfnamefont {E.~S.}\ \bibnamefont
  {Efimenko}}, \bibinfo {author} {\bibfnamefont {A.~V.}\ \bibnamefont
  {Bashinov}}, \bibinfo {author} {\bibfnamefont {A.~A.}\ \bibnamefont
  {Gonoskov}}, \bibinfo {author} {\bibfnamefont {S.~I.}\ \bibnamefont
  {Bastrakov}}, \bibinfo {author} {\bibfnamefont {A.~A.}\ \bibnamefont
  {Muraviev}}, \bibinfo {author} {\bibfnamefont {I.~B.}\ \bibnamefont
  {Meyerov}}, \bibinfo {author} {\bibfnamefont {A.~V.}\ \bibnamefont {Kim}},\
  and\ \bibinfo {author} {\bibfnamefont {A.~M.}\ \bibnamefont {Sergeev}},\
  }\bibfield  {title} {\bibinfo {title} {Laser-driven plasma pinching in e-e+
  cascade},\ }\href {https://doi.org/10.1103/PhysRevE.99.031201} {\bibfield
  {journal} {\bibinfo  {journal} {Physical Review E}\ }\textbf {\bibinfo
  {volume} {99}},\ \bibinfo {pages} {031201} (\bibinfo {year}
  {2019})}\BibitemShut {NoStop}%
\bibitem [{\citenamefont {Efimenko}\ \emph {et~al.}(2018)\citenamefont
  {Efimenko}, \citenamefont {Bashinov}, \citenamefont {Bastrakov},
  \citenamefont {Gonoskov}, \citenamefont {Muraviev}, \citenamefont {Meyerov},
  \citenamefont {Kim},\ and\ \citenamefont {Sergeev}}]{efimenko.sr.2018}%
  \BibitemOpen
  \bibfield  {author} {\bibinfo {author} {\bibfnamefont {E.~S.}\ \bibnamefont
  {Efimenko}}, \bibinfo {author} {\bibfnamefont {A.~V.}\ \bibnamefont
  {Bashinov}}, \bibinfo {author} {\bibfnamefont {S.~I.}\ \bibnamefont
  {Bastrakov}}, \bibinfo {author} {\bibfnamefont {A.~A.}\ \bibnamefont
  {Gonoskov}}, \bibinfo {author} {\bibfnamefont {A.~A.}\ \bibnamefont
  {Muraviev}}, \bibinfo {author} {\bibfnamefont {I.~B.}\ \bibnamefont
  {Meyerov}}, \bibinfo {author} {\bibfnamefont {A.~V.}\ \bibnamefont {Kim}},\
  and\ \bibinfo {author} {\bibfnamefont {A.~M.}\ \bibnamefont {Sergeev}},\
  }\bibfield  {title} {\bibinfo {title} {{Extreme plasma states in
  laser-governed vacuum breakdown}},\ }\href
  {https://doi.org/10.1038/s41598-018-20745-y} {\bibfield  {journal} {\bibinfo
  {journal} {Scientific Reports}\ }\textbf {\bibinfo {volume} {8}},\ \bibinfo
  {pages} {2329} (\bibinfo {year} {2018})}\BibitemShut {NoStop}%
\bibitem [{\citenamefont {Gelfer}\ \emph {et~al.}(2015)\citenamefont {Gelfer},
  \citenamefont {Mironov}, \citenamefont {Fedotov}, \citenamefont {Bashmakov},
  \citenamefont {Nerush}, \citenamefont {Kostyukov},\ and\ \citenamefont
  {Narozhny}}]{gelfer.pra.2015}%
  \BibitemOpen
  \bibfield  {author} {\bibinfo {author} {\bibfnamefont {E.~G.}\ \bibnamefont
  {Gelfer}}, \bibinfo {author} {\bibfnamefont {A.~A.}\ \bibnamefont {Mironov}},
  \bibinfo {author} {\bibfnamefont {A.~M.}\ \bibnamefont {Fedotov}}, \bibinfo
  {author} {\bibfnamefont {V.~F.}\ \bibnamefont {Bashmakov}}, \bibinfo {author}
  {\bibfnamefont {E.~N.}\ \bibnamefont {Nerush}}, \bibinfo {author}
  {\bibfnamefont {I.~Y.}\ \bibnamefont {Kostyukov}},\ and\ \bibinfo {author}
  {\bibfnamefont {N.~B.}\ \bibnamefont {Narozhny}},\ }\bibfield  {title}
  {\bibinfo {title} {Optimized multibeam configuration for observation of qed
  cascades},\ }\href {https://doi.org/10.1103/PhysRevA.92.022113} {\bibfield
  {journal} {\bibinfo  {journal} {Phys. Rev. A}\ }\textbf {\bibinfo {volume}
  {92}},\ \bibinfo {pages} {022113} (\bibinfo {year} {2015})}\BibitemShut
  {NoStop}%
\bibitem [{\citenamefont {Panofsky}(2005)}]{panofsky.2005}%
  \BibitemOpen
  \bibfield  {author} {\bibinfo {author} {\bibfnamefont {W.}~\bibnamefont
  {Panofsky}},\ }\href@noop {} {\emph {\bibinfo {title} {Classical electricity
  and magnetism}}}\ (\bibinfo  {publisher} {Dover Publications},\ \bibinfo
  {address} {Mineola, N.Y},\ \bibinfo {year} {2005})\BibitemShut {NoStop}%
\bibitem [{\citenamefont {Sheppard}\ and\ \citenamefont
  {Larkin}(1994)}]{sheppard.jmo.1994}%
  \BibitemOpen
  \bibfield  {author} {\bibinfo {author} {\bibfnamefont {C.}~\bibnamefont
  {Sheppard}}\ and\ \bibinfo {author} {\bibfnamefont {K.}~\bibnamefont
  {Larkin}},\ }\bibfield  {title} {\bibinfo {title} {Optimal concentration of
  electromagnetic radiation},\ }\href
  {https://doi.org/10.1080/09500349414551421} {\bibfield  {journal} {\bibinfo
  {journal} {Journal of Modern Optics}\ }\textbf {\bibinfo {volume} {41}},\
  \bibinfo {pages} {1495} (\bibinfo {year} {1994})}\BibitemShut {NoStop}%
\bibitem [{\citenamefont {Gonsalves}\ \emph {et~al.}(2019)\citenamefont
  {Gonsalves}, \citenamefont {Nakamura}, \citenamefont {Daniels}, \citenamefont
  {Benedetti}, \citenamefont {Pieronek}, \citenamefont {de~Raadt},
  \citenamefont {Steinke}, \citenamefont {Bin}, \citenamefont {Bulanov},
  \citenamefont {van Tilborg}, \citenamefont {Geddes}, \citenamefont
  {Schroeder}, \citenamefont {T{\'{o}}th}, \citenamefont {Esarey},
  \citenamefont {Swanson}, \citenamefont {Fan-Chiang}, \citenamefont
  {Bagdasarov}, \citenamefont {Bobrova}, \citenamefont {Gasilov}, \citenamefont
  {Korn}, \citenamefont {Sasorov},\ and\ \citenamefont
  {Leemans}}]{gonsalves.prl.2019}%
  \BibitemOpen
  \bibfield  {author} {\bibinfo {author} {\bibfnamefont {A.}~\bibnamefont
  {Gonsalves}}, \bibinfo {author} {\bibfnamefont {K.}~\bibnamefont {Nakamura}},
  \bibinfo {author} {\bibfnamefont {J.}~\bibnamefont {Daniels}}, \bibinfo
  {author} {\bibfnamefont {C.}~\bibnamefont {Benedetti}}, \bibinfo {author}
  {\bibfnamefont {C.}~\bibnamefont {Pieronek}}, \bibinfo {author}
  {\bibfnamefont {T.}~\bibnamefont {de~Raadt}}, \bibinfo {author}
  {\bibfnamefont {S.}~\bibnamefont {Steinke}}, \bibinfo {author} {\bibfnamefont
  {J.}~\bibnamefont {Bin}}, \bibinfo {author} {\bibfnamefont {S.}~\bibnamefont
  {Bulanov}}, \bibinfo {author} {\bibfnamefont {J.}~\bibnamefont {van
  Tilborg}}, \bibinfo {author} {\bibfnamefont {C.}~\bibnamefont {Geddes}},
  \bibinfo {author} {\bibfnamefont {C.}~\bibnamefont {Schroeder}}, \bibinfo
  {author} {\bibfnamefont {C.}~\bibnamefont {T{\'{o}}th}}, \bibinfo {author}
  {\bibfnamefont {E.}~\bibnamefont {Esarey}}, \bibinfo {author} {\bibfnamefont
  {K.}~\bibnamefont {Swanson}}, \bibinfo {author} {\bibfnamefont
  {L.}~\bibnamefont {Fan-Chiang}}, \bibinfo {author} {\bibfnamefont
  {G.}~\bibnamefont {Bagdasarov}}, \bibinfo {author} {\bibfnamefont
  {N.}~\bibnamefont {Bobrova}}, \bibinfo {author} {\bibfnamefont
  {V.}~\bibnamefont {Gasilov}}, \bibinfo {author} {\bibfnamefont
  {G.}~\bibnamefont {Korn}}, \bibinfo {author} {\bibfnamefont {P.}~\bibnamefont
  {Sasorov}},\ and\ \bibinfo {author} {\bibfnamefont {W.}~\bibnamefont
  {Leemans}},\ }\bibfield  {title} {\bibinfo {title} {Petawatt laser guiding
  and electron beam acceleration to 8~{GeV} in a laser-heated capillary
  discharge waveguide},\ }\bibfield  {journal} {\bibinfo  {journal} {Physical
  Review Letters}\ }\textbf {\bibinfo {volume} {122}},\ \href
  {https://doi.org/10.1103/physrevlett.122.084801}
  {10.1103/physrevlett.122.084801} (\bibinfo {year} {2019})\BibitemShut
  {NoStop}%
\bibitem [{\citenamefont {Blackburn}\ \emph {et~al.}(2019)\citenamefont
  {Blackburn}, \citenamefont {Ilderton}, \citenamefont {Marklund},\ and\
  \citenamefont {Ridgers}}]{blackburn.njp.2019}%
  \BibitemOpen
  \bibfield  {author} {\bibinfo {author} {\bibfnamefont {T.~G.}\ \bibnamefont
  {Blackburn}}, \bibinfo {author} {\bibfnamefont {A.}~\bibnamefont {Ilderton}},
  \bibinfo {author} {\bibfnamefont {M.}~\bibnamefont {Marklund}},\ and\
  \bibinfo {author} {\bibfnamefont {C.~P.}\ \bibnamefont {Ridgers}},\
  }\bibfield  {title} {\bibinfo {title} {Reaching supercritical field strengths
  with intense lasers},\ }\href {https://doi.org/10.1088/1367-2630/ab1e0d}
  {\bibfield  {journal} {\bibinfo  {journal} {New Journal of Physics}\ }\textbf
  {\bibinfo {volume} {21}},\ \bibinfo {pages} {053040} (\bibinfo {year}
  {2019})}\BibitemShut {NoStop}%
\bibitem [{\citenamefont {Rivas}\ \emph {et~al.}(2017)\citenamefont {Rivas},
  \citenamefont {Borot}, \citenamefont {Cardenas}, \citenamefont {Marcus},
  \citenamefont {Gu}, \citenamefont {Herrmann}, \citenamefont {Xu},
  \citenamefont {Tan}, \citenamefont {Kormin}, \citenamefont {Ma},
  \citenamefont {Dallari}, \citenamefont {Tsakiris}, \citenamefont
  {F\"{o}ldes}, \citenamefont {w.~Chou}, \citenamefont {Weidman}, \citenamefont
  {Bergues}, \citenamefont {Wittmann}, \citenamefont {Schr\"{o}der},
  \citenamefont {Tzallas}, \citenamefont {Charalambidis}, \citenamefont
  {Razskazovskaya}, \citenamefont {Pervak}, \citenamefont {Krausz},\ and\
  \citenamefont {Veisz}}]{rivas.sr.2017}%
  \BibitemOpen
  \bibfield  {author} {\bibinfo {author} {\bibfnamefont {D.~E.}\ \bibnamefont
  {Rivas}}, \bibinfo {author} {\bibfnamefont {A.}~\bibnamefont {Borot}},
  \bibinfo {author} {\bibfnamefont {D.~E.}\ \bibnamefont {Cardenas}}, \bibinfo
  {author} {\bibfnamefont {G.}~\bibnamefont {Marcus}}, \bibinfo {author}
  {\bibfnamefont {X.}~\bibnamefont {Gu}}, \bibinfo {author} {\bibfnamefont
  {D.}~\bibnamefont {Herrmann}}, \bibinfo {author} {\bibfnamefont
  {J.}~\bibnamefont {Xu}}, \bibinfo {author} {\bibfnamefont {J.}~\bibnamefont
  {Tan}}, \bibinfo {author} {\bibfnamefont {D.}~\bibnamefont {Kormin}},
  \bibinfo {author} {\bibfnamefont {G.}~\bibnamefont {Ma}}, \bibinfo {author}
  {\bibfnamefont {W.}~\bibnamefont {Dallari}}, \bibinfo {author} {\bibfnamefont
  {G.~D.}\ \bibnamefont {Tsakiris}}, \bibinfo {author} {\bibfnamefont {I.~B.}\
  \bibnamefont {F\"{o}ldes}}, \bibinfo {author} {\bibfnamefont
  {S.}~\bibnamefont {w.~Chou}}, \bibinfo {author} {\bibfnamefont
  {M.}~\bibnamefont {Weidman}}, \bibinfo {author} {\bibfnamefont
  {B.}~\bibnamefont {Bergues}}, \bibinfo {author} {\bibfnamefont
  {T.}~\bibnamefont {Wittmann}}, \bibinfo {author} {\bibfnamefont
  {H.}~\bibnamefont {Schr\"{o}der}}, \bibinfo {author} {\bibfnamefont
  {P.}~\bibnamefont {Tzallas}}, \bibinfo {author} {\bibfnamefont
  {D.}~\bibnamefont {Charalambidis}}, \bibinfo {author} {\bibfnamefont
  {O.}~\bibnamefont {Razskazovskaya}}, \bibinfo {author} {\bibfnamefont
  {V.}~\bibnamefont {Pervak}}, \bibinfo {author} {\bibfnamefont
  {F.}~\bibnamefont {Krausz}},\ and\ \bibinfo {author} {\bibfnamefont
  {L.}~\bibnamefont {Veisz}},\ }\bibfield  {title} {\bibinfo {title} {Next
  generation driver for attosecond and laser-plasma physics},\ }\bibfield
  {journal} {\bibinfo  {journal} {Scientific Reports}\ }\textbf {\bibinfo
  {volume} {7}},\ \href {https://doi.org/10.1038/s41598-017-05082-w}
  {10.1038/s41598-017-05082-w} (\bibinfo {year} {2017})\BibitemShut {NoStop}%
\bibitem [{\citenamefont {Gustafson}\ \emph {et~al.}(1969)\citenamefont
  {Gustafson}, \citenamefont {Taran}, \citenamefont {Haus}, \citenamefont
  {Lifsitz},\ and\ \citenamefont {Kelley}}]{gustafson.pr.1969}%
  \BibitemOpen
  \bibfield  {author} {\bibinfo {author} {\bibfnamefont {T.~K.}\ \bibnamefont
  {Gustafson}}, \bibinfo {author} {\bibfnamefont {J.~P.}\ \bibnamefont
  {Taran}}, \bibinfo {author} {\bibfnamefont {H.~A.}\ \bibnamefont {Haus}},
  \bibinfo {author} {\bibfnamefont {J.~R.}\ \bibnamefont {Lifsitz}},\ and\
  \bibinfo {author} {\bibfnamefont {P.~L.}\ \bibnamefont {Kelley}},\ }\bibfield
   {title} {\bibinfo {title} {Self-modulation, self-steepening, and spectral
  development of light in small-scale trapped filaments},\ }\href
  {https://doi.org/10.1103/physrev.177.306} {\bibfield  {journal} {\bibinfo
  {journal} {Physical Review}\ }\textbf {\bibinfo {volume} {177}},\ \bibinfo
  {pages} {306} (\bibinfo {year} {1969})}\BibitemShut {NoStop}%
\bibitem [{\citenamefont {Kaw}(1970)}]{kaw.pf.1970}%
  \BibitemOpen
  \bibfield  {author} {\bibinfo {author} {\bibfnamefont {P.}~\bibnamefont
  {Kaw}},\ }\bibfield  {title} {\bibinfo {title} {Relativistic nonlinear
  propagation of laser beams in cold overdense plasmas},\ }\href
  {https://doi.org/10.1063/1.1692942} {\bibfield  {journal} {\bibinfo
  {journal} {Physics of Fluids}\ }\textbf {\bibinfo {volume} {13}},\ \bibinfo
  {pages} {472} (\bibinfo {year} {1970})}\BibitemShut {NoStop}%
\bibitem [{\citenamefont {Gonoskov}\ \emph {et~al.}(2009)\citenamefont
  {Gonoskov}, \citenamefont {Korzhimanov}, \citenamefont {Eremin},
  \citenamefont {Kim},\ and\ \citenamefont {Sergeev}}]{gonoskov.prl.2009}%
  \BibitemOpen
  \bibfield  {author} {\bibinfo {author} {\bibfnamefont {A.~A.}\ \bibnamefont
  {Gonoskov}}, \bibinfo {author} {\bibfnamefont {A.~V.}\ \bibnamefont
  {Korzhimanov}}, \bibinfo {author} {\bibfnamefont {V.~I.}\ \bibnamefont
  {Eremin}}, \bibinfo {author} {\bibfnamefont {A.~V.}\ \bibnamefont {Kim}},\
  and\ \bibinfo {author} {\bibfnamefont {A.~M.}\ \bibnamefont {Sergeev}},\
  }\bibfield  {title} {\bibinfo {title} {Multicascade proton acceleration by a
  superintense laser pulse in the regime of relativistically induced slab
  transparency},\ }\bibfield  {journal} {\bibinfo  {journal} {Physical Review
  Letters}\ }\textbf {\bibinfo {volume} {102}},\ \href
  {https://doi.org/10.1103/physrevlett.102.184801}
  {10.1103/physrevlett.102.184801} (\bibinfo {year} {2009})\BibitemShut
  {NoStop}%
\bibitem [{\citenamefont {Reed}\ \emph {et~al.}(2009)\citenamefont {Reed},
  \citenamefont {Matsuoka}, \citenamefont {Bulanov}, \citenamefont {Tampo},
  \citenamefont {Chvykov}, \citenamefont {Kalintchenko}, \citenamefont
  {Rousseau}, \citenamefont {Yanovsky}, \citenamefont {Kodama}, \citenamefont
  {Litzenberg}, \citenamefont {Krushelnick},\ and\ \citenamefont
  {Maksimchuk}}]{reed.apl.2009}%
  \BibitemOpen
  \bibfield  {author} {\bibinfo {author} {\bibfnamefont {S.~A.}\ \bibnamefont
  {Reed}}, \bibinfo {author} {\bibfnamefont {T.}~\bibnamefont {Matsuoka}},
  \bibinfo {author} {\bibfnamefont {S.}~\bibnamefont {Bulanov}}, \bibinfo
  {author} {\bibfnamefont {M.}~\bibnamefont {Tampo}}, \bibinfo {author}
  {\bibfnamefont {V.}~\bibnamefont {Chvykov}}, \bibinfo {author} {\bibfnamefont
  {G.}~\bibnamefont {Kalintchenko}}, \bibinfo {author} {\bibfnamefont
  {P.}~\bibnamefont {Rousseau}}, \bibinfo {author} {\bibfnamefont
  {V.}~\bibnamefont {Yanovsky}}, \bibinfo {author} {\bibfnamefont
  {R.}~\bibnamefont {Kodama}}, \bibinfo {author} {\bibfnamefont {D.~W.}\
  \bibnamefont {Litzenberg}}, \bibinfo {author} {\bibfnamefont
  {K.}~\bibnamefont {Krushelnick}},\ and\ \bibinfo {author} {\bibfnamefont
  {A.}~\bibnamefont {Maksimchuk}},\ }\bibfield  {title} {\bibinfo {title}
  {Relativistic plasma shutter for ultraintense laser pulses},\ }\href
  {https://doi.org/10.1063/1.3139860} {\bibfield  {journal} {\bibinfo
  {journal} {Applied Physics Letters}\ }\textbf {\bibinfo {volume} {94}},\
  \bibinfo {pages} {201117} (\bibinfo {year} {2009})}\BibitemShut {NoStop}%
\bibitem [{\citenamefont {Wang}\ \emph {et~al.}(2011)\citenamefont {Wang},
  \citenamefont {Lin}, \citenamefont {Sheng}, \citenamefont {Liu},
  \citenamefont {Zhao}, \citenamefont {Guo}, \citenamefont {Lu}, \citenamefont
  {He}, \citenamefont {Chen},\ and\ \citenamefont {Yan}}]{wang.prl.2011}%
  \BibitemOpen
  \bibfield  {author} {\bibinfo {author} {\bibfnamefont {H.~Y.}\ \bibnamefont
  {Wang}}, \bibinfo {author} {\bibfnamefont {C.}~\bibnamefont {Lin}}, \bibinfo
  {author} {\bibfnamefont {Z.~M.}\ \bibnamefont {Sheng}}, \bibinfo {author}
  {\bibfnamefont {B.}~\bibnamefont {Liu}}, \bibinfo {author} {\bibfnamefont
  {S.}~\bibnamefont {Zhao}}, \bibinfo {author} {\bibfnamefont {Z.~Y.}\
  \bibnamefont {Guo}}, \bibinfo {author} {\bibfnamefont {Y.~R.}\ \bibnamefont
  {Lu}}, \bibinfo {author} {\bibfnamefont {X.~T.}\ \bibnamefont {He}}, \bibinfo
  {author} {\bibfnamefont {J.~E.}\ \bibnamefont {Chen}},\ and\ \bibinfo
  {author} {\bibfnamefont {X.~Q.}\ \bibnamefont {Yan}},\ }\bibfield  {title}
  {\bibinfo {title} {Laser shaping of a relativistic intense, short gaussian
  pulse by a plasma lens},\ }\bibfield  {journal} {\bibinfo  {journal}
  {Physical Review Letters}\ }\textbf {\bibinfo {volume} {107}},\ \href
  {https://doi.org/10.1103/physrevlett.107.265002}
  {10.1103/physrevlett.107.265002} (\bibinfo {year} {2011})\BibitemShut
  {NoStop}%
\end{thebibliography}%

\end{document}